# A framework for scale-sensitive, spatially explicit accuracy assessment of binary built-up surface layers

**Johannes H. Uhl[1,2,3], Stefan Leyk[1,3]**

[1]University of Colorado Boulder, Department of Geography, 260 UCB, Boulder, CO-80309, USA.
[2]University of Colorado Boulder, Cooperative Institute for Research in Environmental Sciences (CIRES) 216 UCB, Boulder, CO-80309, USA.
[3]University of Colorado Boulder, Institute of Behavioral Science, 483 UCB, Boulder, CO-80309, USA.
Email: {Johannes.Uhl;Stefan.Leyk}@colorado.edu
Corresponding author: Johannes H. Uhl (Johannes.Uhl@colorado.edu)

**Abstract:** To better understand the dynamics of human settlements, thorough knowledge of the uncertainty in geospatial built-up surface datasets is critical. While frameworks for localized accuracy assessments of categorical gridded data have been proposed to account for the spatial non-stationarity of classification accuracy, such approaches have not been applied to (binary) built-up land data. Such data differs from other data such as land cover data, due to considerable variations of built-up surface density across the rural-urban continuum resulting in switches of class imbalance, causing sparsely populated confusion matrices based on small underlying sample sizes. In this paper, we aim to fill this gap by testing common agreement measures for their suitability and plausibility to measure the localized accuracy of built-up surface data. We examine the sensitivity of localized accuracy to the assessment support, as well as to the unit of analysis, and analyze the relationships between local accuracy and density / structure-related properties of built-up areas, across rural-urban trajectories and over time. Our experiments are based on the multi-temporal Global Human Settlement Layer (GHSL) and a reference database for the state of Massachusetts (USA). We find strong variation of suitability among commonly used agreement measures, and varying levels of sensitivity to the assessment support. We then apply our framework to assess localized GHSL data accuracy over time from 1975 to 2014. Besides increasing accuracy along the rural-urban gradient, we find that accuracy generally increases over time, mainly driven by peri-urban densification processes in our study area. Moreover, we find that localized densification measures derived from the GHSL tend to overestimate peri-urban densification processes that occurred between 1975 and 2014, due to higher levels of omission errors in the GHSL epoch 1975.

## 1. Introduction

Understanding the regional and local patterns of human settlements on Earth requires not only reliable geospatial data at sufficiently high spatial resolution, but also thorough knowledge about the uncertainty in the data used to analyze settlement processes (e.g., suburbanization and conurbation processes), including the spatial (and temporal) variability of the uncertainty inherent in built-up land data. Ignoring or oversimplifying the uncertainty of such data may seriously bias the interpretation of analytical results, and thus, frameworks for the accuracy assessment of such data products are required to take into account the peculiarities of built-up land data. This includes the accuracy estimation at suitable analytical extents and within meaningful spatial strata. Moreover, local processes of human settlement (e.g., densification, expansion, or infilling processes) can only be modeled and understood objectively if the underlying local uncertainty structure is known.

Uncertainty in geospatial categorical data such as remote-sensing derived land cover data is often quantified by statistical measures obtained through accuracy assessments that are based on map comparison techniques. In such assessments the examined data are compared to an independently compiled reference dataset of presumed higher accuracy (FGDC 1998). Common map comparison approaches include the use of confusion matrices to derive accuracy metrics that quantify the agreement between the test data and reference data within the study area (Fielding and Bell 1997). In a traditional accuracy assessment, a global accuracy measure is computed for the whole study area ignoring spatial variation of the level of agreement between the two data sources (Foody 2007). In recent years, geospatial research has established an improved understanding of uncertainty in spatial data as a spatially varying phenomenon (e.g., Leyk and Zimmermann 2004, Foody 2007, Wickham et al. 2018). This spatial variation can be driven by differences in quality of underlying source data between rural and urban regions, ambiguous spectral responses of different land cover categories, or obstructions due to clouds, to mention some examples.

Based on this recent research, it is known that overly aggregated accuracy measures might misrepresent the inherent uncertainty in the data under test and ignore its spatial structure. Furthermore, it has been shown that classification accuracy metrics can be sensitive to the sample size (e.g., Sim and Wright 2005, Bujang and Baharum 2017, Champagne et al. 2014) and severely biased if the proportional sizes of individual classes are heavily imbalanced (see Rosenfield and Melley 1980, Wickham et al. 2010, Akosa 2017, Shao et al. 2019, Radoux et al 2020, Stehman and Wickham 2020). To reduce these effects, different approaches have been proposed including stratified sampling (e.g., Congalton 1991), spatially constrained (localized) confusion matrices (Leyk and Zimmermann 2004, Foody 2007), predictive uncertainty modelling using ancillary variables (e.g., Smith et al. 2003, Leyk and Zimmermann 2007, van Oort et al. 2004, Zhang and Mei 2016, Wickham et al. 2018, Mei et al. 2019) and spatial / geostatistical interpolation methods (Steele et al. 1998, Kyriakidis and Dungan 2001, Comber et al. 2012, Tsutsumida and Comber 2015). Localized accuracy assessments typically involve the creation of continuous accuracy surfaces and are sometimes referred to as spatially explicit accuracy assessments (Löw et al. 2013, Khatami et al. 2017, Waldner et al. 2017, Mitchell et al. 2018).

According to Foody (2002), accuracy estimates may vary considerably for different analytical scales, and depend strongly on the sample used to establish the confusion matrix, which ideally is representative for the conditions found within the study area (see also Stehman and Foody 2019). Accordingly, different efforts have proposed and applied accuracy assessment frameworks using different analytical units (Pontius 2002, Pontius and Suedmeyer 2004, Pontius et al. 2004, Pontius and Cheuk 2006, Pontius et al. 2008b, Pontius et al. 2011, Stehman and Wickham 2011, Zhu et al. 2013, Yan et al. 2014, Ye et al. 2018, Marconcini et al. 2020a), for different sample sizes (e.g., Congalton 1988, Hashemian et al. 2004, Foody 2009) but also across different geographic extents (Wardlow and Callahan 2014, Ariza-López et al. 2018), and different levels of semantic aggregation (Pontius and Malizia 2004). The geographic extent (sometimes called geographic scale, cf. Smith 2000) used to draw a sample of analytical units to establish the confusion matrix is the spatial support, or assessment unit (Stehman 2009), and will be called ***assessment support*** in this work.

The sensitivity of a spatial variable to the size and shape of an imposed zoning unit used for aggregation is a well-known phenomenon in geographic information science and the social sciences (i.e., the modifiable areal unit problem, MAUP; Openshaw 1984, see Nelson and Brewer 2017 for a recent in-depth study). Hence, it is particularly surprising that only few studies have analyzed the sensitivity of accuracy measures to their constraining geometry or assessment support, considering that the elements of the confusion matrix proposed for a given areal extent consist of the sums of agreement-disagreement combinations within that extent and thus, can be conceptualized as a spatially aggregated geographic variable that propagates the inherent uncertainties of the selected analytical scales.

With recent technological advances in geospatial data acquisition, processing, cloud-based dissemination and analysis infrastructure, there is an increasing amount of novel geospatial datasets available, measuring the spatio-temporal distribution of human settlements and land cover in general, over large extents and at unprecedented spatial granularity. These datasets include the different built-up surface layers from the Global Human Settlement Layer project (GHSL, Pesaresi et al. 2013, Corbane et al. 2019a, Corbane et al. 2019b, Corbane et al. 2022), Global Urban Footprint (Esch et al. 2013), High-Resolution Settlement Layer (Facebook Connectivity Lab and CIESIN 2016), and the World Settlement Footprint (Marconcini et al. 2020a and 2020b, as well as the FROM-GLC10 (Gong et al. 2019), and the Global artificial impervious areas product (GAIA, Gong et al. 2020). While such datasets greatly facilitate the study of urbanization, human-natural systems and related geographic-environmental processes at unseen levels of detail, little research has been done on the accuracy of such datasets and how accuracy trajectories can be characterized across the rural-urban continuum, often due to the lack of reliable reference data over sufficiently large spatial (and temporal) extents. For example, previous work has revealed varying levels of accuracy among different settlement datasets (Klotz et al. 2016), increasing accuracy levels over time in case of the multi-temporal Global Human Settlement Layer (Leyk et al. 2018), and increases in accuracy from rural towards urban areas (Uhl and Leyk 2017, Uhl et al. 2018, Liu et al. 2020). However, these general trends are based on coarse, regional stratification of the studied area and thus, possibly neglect local accuracy variations.

High-resolution built-up land data, discriminating between built-up and not built-up land in a binary fashion, exhibit some significant differences compared to multi-class land use / land cover (LULC) data, that is, they can be severely imbalanced, and this imbalance can shift between rural and urban areas. Furthermore, measures derived from localized confusion matrices can be void due to zero instances in one of the matrix fields. Thus, a framework for localized accuracy assessment of built-up land data needs to account for extreme, bi-directional class imbalance, as well as small sample sizes underlying a spatially constrained confusion matrix, and the absence of instances of one or more confusion matrix elements. This study has the goal to develop such a framework guided by the following research questions:

- Are commonly used accuracy measures suitable for assessing the local accuracy of binary, gridded built-up surface datasets?
- How does local accuracy relate to underlying density of built-up area and population density assuming a constant occupancy rate in the study area, and how generalizable are these relationships across the rural-urban continuum, and over time?
- How does the assessment support and analytical unit influence local accuracy estimates and their trends across the rural-urban continuum?

Thus, this study has four contributions: (a) we identify suitable accuracy measures for localized uncertainty assessment of built-up land data, (b) we reveal novel, fine-grained insights of the local, spatio-temporal uncertainty inherent in the multitemporal, Landsat-based Global Human Settlement Layer, and (c) we assess the scale-dependency of localized accuracy measures. To shed light on these questions, we analyzed the mathematical definition and behavior of commonly used accuracy and agreement measures with respect to small sample sizes and sparsely populated confusion matrices, and we conducted an exemplary, spatially explicit accuracy assessment of built-up area derived from the GHSL against a large reference database derived from cadastral parcel and building footprint records. We generated large amounts (N>100,000,000) of spatially constrained confusion matrices, using (a) external enumeration boundaries to define zones, and (b) moving focal windows as constraining geometry, both at various levels of spatial granularity. We computed a variety of commonly used accuracy measures for these zonal and focal constraining regions, to assess their sensitivity to the assessment support, and examined relationships between these local accuracy measures and structure / density of built-up area, as well as population density. Finally, we applied our framework to the multi-temporal settlement data from the GHSL, and assessed temporal trajectories of localized accuracy across space and along the rural-urban continuum. Herein, we will use the term "built-up density" when referring to the density of built-up surface within a given areal reference unit. Moreover, we will use the term "local / localized accuracy" for focal and zonal accuracy estimates, describing the data accuracy within a local spatial unit. The term "accuracy" refers to both, estimates of thematic and quantity agreement (see Section 2.2.3).

## 2. Data and methods

This study consists of two major analytical parts: First, we analyzed accuracy measures of GHSL-derived built-up areas within spatial units defined by zoning data derived from administrative boundaries and U.S. census enumeration units of various granularities (i.e., *zonal* accuracy estimates). Second, we assessed accuracy measures within moving windows of varying size (i.e., *focal* accuracy estimates). The former allowed for examining relationships of zonal accuracy estimates to population density, whereas the latter allowed for generating a continuous space of assessment support, independently of underlying administrative zones or population distributions. In both cases, we employed a highly accurate reference database of built-up areas derived from cadastral parcel and building footprint data and analyzed these localized accuracy measures in various ways for the state of Massachusetts, USA, which extends across an area of over 27,000 km² and contains highly urbanized regions, such as Boston, but also extensive rural areas.

### 2.1 Data

This study is based on binary built-up / not built-up raster layers extracted from the GHSL (Figure 1a,b). More specifically, we employed built-up areas in 1975, 1990, 2000 and 2014 extracted from the GHSL Landsat edition (GHS-BUILT R2018A, Florczyk et al. 2019, file name: GHS_BUILT_LDSMT_GLOBE_R2018A_3857_30_V2_0). While finer-grained, contemporary built-up land depictions have been released in the GHSL effort (e.g. Corbane et al. 2021), the GHS-BUILT R2018A is, to date, the most recent, and fine-grained global settlement dataset covering such a long time period. The GHSL estimates the presence and distribution of human settlements on the planet at a spatial resolution of 30 meters and for different points in time (1975, 1990, 2000, and 2014), based on multi-temporal Landsat data and a machine learning approach (Pesaresi et al. 2015, 2016). We used the GHS-BUILT R2018A, as it extends farthest back in time among the multi-temporal global built-up surface datasets (i.e., to 1975, as opposed to the WSF-evolution data product dating back to 1985, Marconcini et al. 2020b). Moreover, the GHSL dataset has been used for different data production efforts, such as the GHS-POP population dataset or the GHS-SMOD rural-urban classification (Florczyk et al. 2019), The GHS-BUILT R2018A or derived products have been used in a wide range of scientific studies (see Ehrlich et al. 2021 for an overview). Thus, understanding the uncertainty in this data product enables a more reflected use of the data or derived datasets in applied studies.

The extracted built-up presence surfaces represent the data under test, and were compared to a reference database of multi-temporal built-up areas in the U.S. that has been created by the authors through integrating publicly available cadastral, tax assessment and building footprint data and allows for accuracy assessments of built-up land data at fine spatial resolution, covering over 30 U.S. counties (i.e., >40,000km², more than 6,000,000 cadastral parcels). Parcel geometries which include built-year information were spatially refined to the extent of building outlines and rasterized using the GHSL grid properties (Figure 1c,d). This multi-temporal reference database has been applied for validation purposes in previous work (see Uhl and Leyk 2017, Leyk et al. 2018, Uhl et al. 2018, Leyk and Uhl 2018, Uhl et al. 2021). We call this database the ***Multi-Temporal Building Footprint dataset*** (MTBF-33) as it covers 33 U.S. counties and made this database publicly available (Uhl & Leyk 2022). This valuable data source can be used to create unique snapshots of built-up land suitable as reference surfaces for developed or built-up land classes at arbitrary points in time since 1900 and earlier. We assessed the plausibility of this integrated data product by cross-comparing building and parcel information and excluded discrepant areas from the analysis (e.g., parcels without associated building footprint but indicating the presence of a building, making up approximately 16% of the study area), which increases the reliability of the reference data (see Leyk et al. 2018 for details). The state of Massachusetts represents the largest contiguous area covered in MTBF-33 and thus, is used as study area herein.

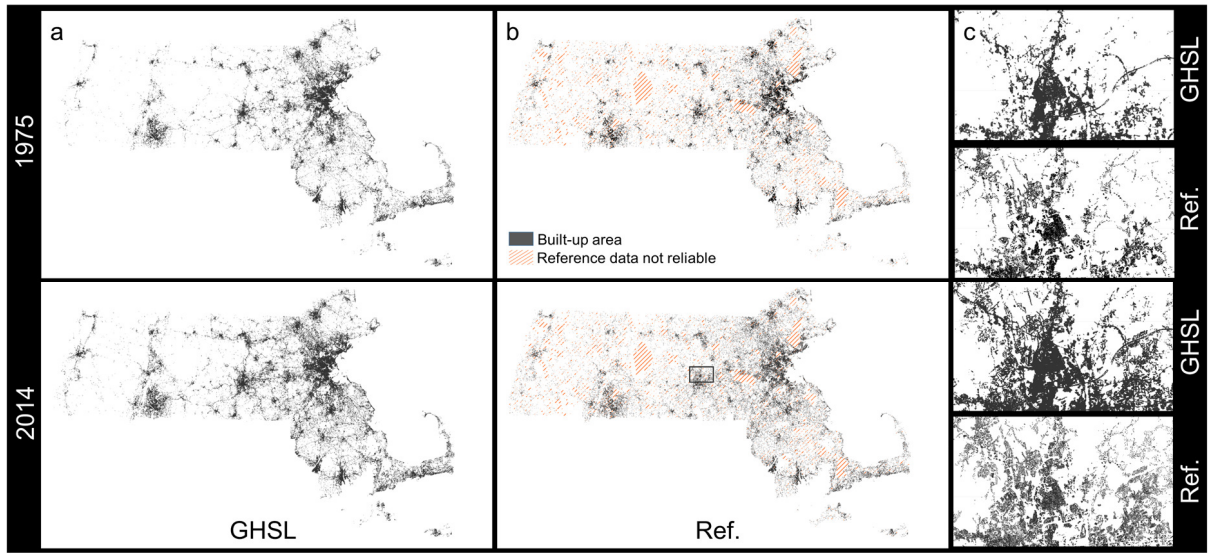

**Figure 1. Data used in this study: Built-up areas (a) from the GHS_BUILT_LDSMT_GLOBE_R2018A product in 1975 and 2014, and (b) from the reference database, at a spatial resolution of 30mx30m, for Massachusetts (USA). Panel (c) shows the data for a part of the city of Worcester, Massachusetts.**

We derived the zoning data from administrative boundaries (i.e., state, county and township boundaries; MassGIS 2016) and U.S. census enumeration units, (i.e., census tracts, block groups and blocks; U.S. Census Bureau 2016). Census tracts generally have a population size between 1,200 and 8,000 people, block groups contain between 600 and 3,000 people and census blocks represent single city blocks in urban areas, and may encompass large areas in rural regions (U.S. Census Bureau 2017). In 2010, the state of Massachusetts contains 14 counties, 351 townships, 1,475 census tracts, 4,982 block groups, and 157,508 census blocks. The delineation of census enumeration boundaries is heavily influenced by the underlying spatial population distribution, and it can be expected that large-scale spatio-temporal patterns of population are related to those of built-up area. Therefore, using census enumeration units is an inherently meaningful way to constrain the confusion matrices spatially for substantive evaluation of underlying accuracy associations. The levels of granularity of these spatial units constitute different levels of *assessment support*. The use of census data from 2010 and GHSL built-up areas from 2014 ensured that temporal discrepancies between zoning and test data were kept to a minimum.

### 2.2 Methods

In a first step, we projected and rasterized the built-up areas derived from the reference database in 2014 (i.e., polygonal vector data) to the spatial reference system and spatial resolution used in GHSL version 2018. During this process, the definition of the abstract class "built-up area" (i.e., a grid cell is considered built-up if at least one built-up structure overlaps the grid cell; see Pesaresi et al. 2016) was applied to the reference data to ensure spatial and thematic compatibility (Figure 2a).

#### 2.2.1 Generating zonal accuracy measures

First, we generated agreement category surfaces, i.e., encoding true positives (TP), true negatives (TN), false positives (FP), and false negatives (FN) based on map comparison (i.e., pixel-wise agreement / disagreement) between built-up areas in 2014 derived from GHSL and the rasterized reference data (Figure 2b), using an exhaustive sampling scheme, excluding grid cells within parcels that are considered unreliable. These surfaces use one-hot encoding (e.g., TP = 1, other cells = 0). For each agreement category, we computed the zonal sums of the respective categories within each (vector) zoning geometry at all six assessment support levels (i.e., state, county, township, tract, block group, block), yielding the confusion matrix for each individual zoning geometry (Figure 2c). We calculated a range of accuracy measures (Section 2.2.3), derived from these confusion matrices and appended them as attributes to the respective zoning geometries. We linked each of the >150,000 census blocks to all zoning geometries that spatially contained the respective census block, in order to establish links between all zoning geometries across the assessment support domain. This method allows for extracting accuracy measures at each individual zoning geometry defining the assessment support, , as well as the accuracy trajectories for a given location across all support levels for visualization and analysis (see Section 3.4).

#### 2.2.2 Generating focal accuracy measures

The second data processing effort conducted in this study yielded a set of surfaces of localized accuracy measures within focal (moving) windows, of varying size, and thus, independent from external zoning data (Figure 2d). More specifically, we used quadratic focal windows of size *s*x*s*, with *s* ϵ (1km, 2.5km, 5km, 10km), representing four levels of focal assessment support. In a first step, for each focal support level, we computed the focal sum of the instances of each agreement type (TP, TN, FP, FN, see Figure 2e). For example, the focal sum of TP instances for support *s*=1000m (i.e., $TP_{1000}$) represents the TP elements of the corresponding localized confusion matrices $CM_{1000}$. We stacked these four surfaces into a 4-band composite, representing a spatialized version of localized confusion matrices.. We generated such a confusion matrix composite for each of the four support levels. The TP, FP, and FN bands of these composites are shown in Figure 2f using RGB color-coding, exemplarily for *s*=1km and *s*=2.5 km.

For each of the four levels of focal assessment support, we drew a stratified random subsample of N=1,000,000 locations from the >6.6 million grid cells within Massachusetts that have at least one GHSL or reference built-up instance within their focal neighborhood. In order to obtain a representative sample across the rural-urban continuum covering both GHSL and reference data, we stratified the data by deciles of reference built-up area density (i.e., 100,000 locations per decile stratum). All subsequent computations are based on these compositional data structures, allowing for efficient retrieval of localized confusion matrix components at any location and support level, and the fast computation of accuracy surfaces for a range of accuracy measures (see Figure 2g for an example, see also Section 2.2.3) and built-up area density surfaces (Figure 2g, Section 2.2.4).

### 2.2.3 Agreement measures

The agreement measures examined herein are based on a binary contingency table, representing the confusion matrix CM:"

$$CM = \begin{bmatrix} TN & FP \\ FN & TP \end{bmatrix} \tag{1}$$

with TN: true negatives, FN: false negatives, FP: false positives, and TP: true positives as counts resulting from cross-tabulation of the reference and test data records, where "positive" refers to the entities of interest. Then the overall accuracy, or percentage of correctly classified (PCC) is defined as

$$PCC = \frac{TP + TN}{n} \tag{2}$$

where n is the sum of all elements of CM (Michie et al. 1994). Producer's accuracy (PA, also known as recall, sensitivity, or true positive rate) indicates the probability of a reference element being classified correctly, and is complementary to the omission error OE (error of exclusion, or type II error), whereas User's accuracy (UA, also known as precision) indicates the probability of a classified object being correct, and is complementary to the commission error CE (error of inclusion, or type I error) (Story and Congalton 1986):

$$PA = recall = \frac{TP}{TP + FN} = 1 - OE \tag{3}$$

and

$$UA = precision = \frac{TP}{TP + FP} = 1 - CE \tag{4}$$

Note that in the remainder of this analysis, we use the terms "precision" and "recall". The F-measure is defined as the harmonic mean of precision and recall (Rijsbergen 1974):

$$F = \frac{2 \cdot TP}{2 \cdot TP + FP + FN} = \frac{2 \cdot precision \cdot recall}{precision + recall} \tag{5}$$

and represents a specific case of the generalized $F_\beta$ measure for β=1 (Maratea et al. 2014). The $F_\beta$ measure is defined as:

$$F_\beta = (1 + \beta^2) \cdot \frac{precision \cdot recall}{(\beta^2 \cdot precision) + recall} \tag{6}$$

and allows for assigning higher weights to precision (0 < β < 1) or to recall (β > 1) and is particularly useful to evaluate binary classification scenarios when precision or recall should be emphasized, e.g., in the case of heavily imbalanced data. Some commonly used $F_\beta$ measures are the $F_2$ score (i.e., β=2, favoring recall), and the $F_{0.5}$ score, (i.e., β=0.5, favoring precision) (Rijsbergen 1979).

The geometric mean (G-mean, Kubat and Matwin 1997) is defined as the geometric mean of specificity (i.e., the recall of the negative class) and sensitivity (i.e., recall of the positive class):

$$G = \sqrt{\frac{TN}{TN + FP} \cdot \frac{TP}{FN + TP}} \tag{7}$$

Maratea et al. (2014) combine the concepts of the $F_\beta$ measure and the G-mean and developed the adjusted F-measure ($F_{Adj}$), which is defined as:

$$F_{ADJ} = \sqrt{F_2 \cdot \text{inv}(F_{0.5})} \tag{8}$$

and represents the geometric mean of the $F_2$ measure and inv($F_{0.5}$), where inv($F_{0.5}$) denominates the $F_{0.5}$ measure after inverting the positive and negative class labels, in order to account for the class imbalance bias. Moreover, the Jaccard Index (Jaccard 1902) sometimes referred to as "figure of merit" (e.g., Pontius et al. 2008a), or Intersection-over-Union (IoU) is defined as:

$$IoU = \frac{TP}{TP + FP + FN} \tag{9}$$

Cohen's Kappa index (Cohen 1960) in case of a binary classification is defined as

$$\kappa = \frac{p_0 - p_c}{1 - p_c} \tag{10}$$

with $p_0$ being the observed agreement corresponding to PCC and $p_c$ being chance agreement, estimated as:

$$p_c = \left(\frac{TP + FN}{n}\right)\left(\frac{TP + FP}{n}\right) + \left(\frac{TN + FN}{n}\right)\left(\frac{TN + FP}{n}\right) \tag{11}$$

Moreover, Matthews Correlation Coefficient (MCC, Matthews 1975), defined as:

$$\text{MCC} = \frac{TP \cdot TN - FP \cdot FN}{\sqrt{(TP+FP) \cdot (TP+FN) \cdot (TN+FP) \cdot (TN+FN)}} \tag{12}$$

is increasingly used as an accuracy measure in land cover classifications (e.g., Herfort et al. 2019, Longépé et al. 2019, Vasilakos et al. 2020). Finally, the Normalized Mutual Information score (NMI, Forbes 1995) is obtained based on the entropy $H$ of the predicted class labels $p$ the entropy of the reference class labels $r$ and the entropy of both reference and predicted class labels as:

$$NMI = 1 - \frac{H(r,p) - H(p)}{H(r)} \tag{13}$$

which equals in the case of a binary classification problem to:

$$NMI = 1 - \frac{-TP\ln(TP) - FP\ln(FP) - FN\ln(FN) - TN\ln(TN) + (TP+FP)\ln(TP+FP) + (FN+TN)\ln(FN+TN)}{n\ln(n) - [(TP+FN)\ln(TP+FN) + (FP+TN)\ln(FP+TN)]} \tag{14}$$

Herein, we divided uncertainty into thematic agreement and quantity agreement. A similar separation has been proposed by Pontius and Millones (2011) and has proven to provide interesting insights into model uncertainty (e.g., Pickard et al. 2017), but also into data uncertainty while reducing influences of spatial offsets between test and reference data (see Section 2.2.5).

The quantity agreement measures used herein are the absolute error (AE), obtained as:

$$AE = (TP + FP) - (TP + FN) = FP - FN \tag{15}$$

and the relative error (RE), which is the AE in relation to the built-up quantity reported in the reference data, is calculated as:

$$RE = \frac{AE}{(TP + FN)} = \frac{(FP - FN)}{(TP + FN)} \tag{16}$$

Moreover, we separate the absolute error (Equation 15) into overestimation (OE) and underestimation (UE) components as follows:

$$OE = \begin{cases} AE, & AE > 0 \\ 0, & AE \leq 0 \end{cases} \tag{17}$$

$$UE = \begin{cases} 0, & AE > 0 \\ abs(AE), & AE \leq 0 \end{cases} \tag{18}$$

This will allow for a statistical analysis of the relationships of over- and underestimation components across the rural-urban continuum (Section 2.2.5).

At this point, it is worth noting that, despite being widely used for classification and map accuracy assessments, several of the presented accuracy and agreement measures have been subject to criticisms regarding their suitability for unbiased quantification of classification accuracy. For example, Pontius and Millones (2011) as well as Foody (2020) discourage the community from using the Kappa index, and Shao et al. (2019) and Stehman and Wickham (2020) point out that PCC may be severely biased in case of class imbalance. Conversely, the F-measure and G-mean are known for being less sensitive to imbalance effects (Fawcett 2006), and the MCC has recently been recommended to be preferable over Kappa (Delgado and Tibau 2019), and over the F-measure (Chicco and Jurman 2020), see also Luque et al. (2019). Despite these criticisms, these metrics have been widely used for map comparison and for the evaluation of (binary) classification problems (e.g., Kappa). We include them into our suitability analysis (Section 3.1) to raise further awareness of the potential bias in these metrics (i.e., Kappa and PCC). See Table A1 for an overview of these metrics.

### 2.2.4    Modelling the rural-urban continuum

Quantitative modelling of the rural-urban continuum, i.e., the gradual transition between highly populated urban areas to sparsely populated rural places, represents an important analytical component of this work. While there are numerous global and national data products enabling the high-resolution modelling of the rural-urban continuum based on a variety of input data (e.g., Waldorf and Kim 2018, Florczyk et al. 2019), these datasets use spatial units that are not directly compatible with the assessment support provided by the described localized confusion matrices. Thus, we stratified the study area by varying levels of development intensity, modelled by the built-up density found in the reference data as well as in the GHSL, allowing for stratification across the rural-urban continuum, consistent to the assessment support of the localized accuracy estimates. The reference built-up density (in %) at a given location and for a (quadratic) assessment support $s$ (in m) can be derived from the reference built-up grid cell counts in the confusion matrix composite directly as:

$$BUDENS_{REF,s}[\%] = 100 \cdot 30^2 \cdot \frac{(TP + FN)}{s^2} \tag{19}$$

The GHSL-based built-up density is obtained as:

$$BUDENS_{GHSL,s}[\%] = 100 \cdot 30^2 \cdot \frac{(TP + FP)}{s^2} \tag{20}$$

An example of the resulting focal built-up density surfaces is shown in Figure 2g.
Moreover, we calculated selected landscape metrics quantifying the segregation of the built-up areas. These landscape metrics include the number of built-up patches (NP) and the Largest Patch Index (LPI) (McGarigal et al. 2012). This is motivated by previous work suggesting that particularly the size of patches affects the classification accuracy (Smith et al. 2002 and 2003, Klotz et al. 2016, Mück et al. 2017). We calculated the NP and LPI and assigned them to grids of 30m×30m, consistent to the focal accuracy and density surfaces, for each level of assessment support (Figure 2h).

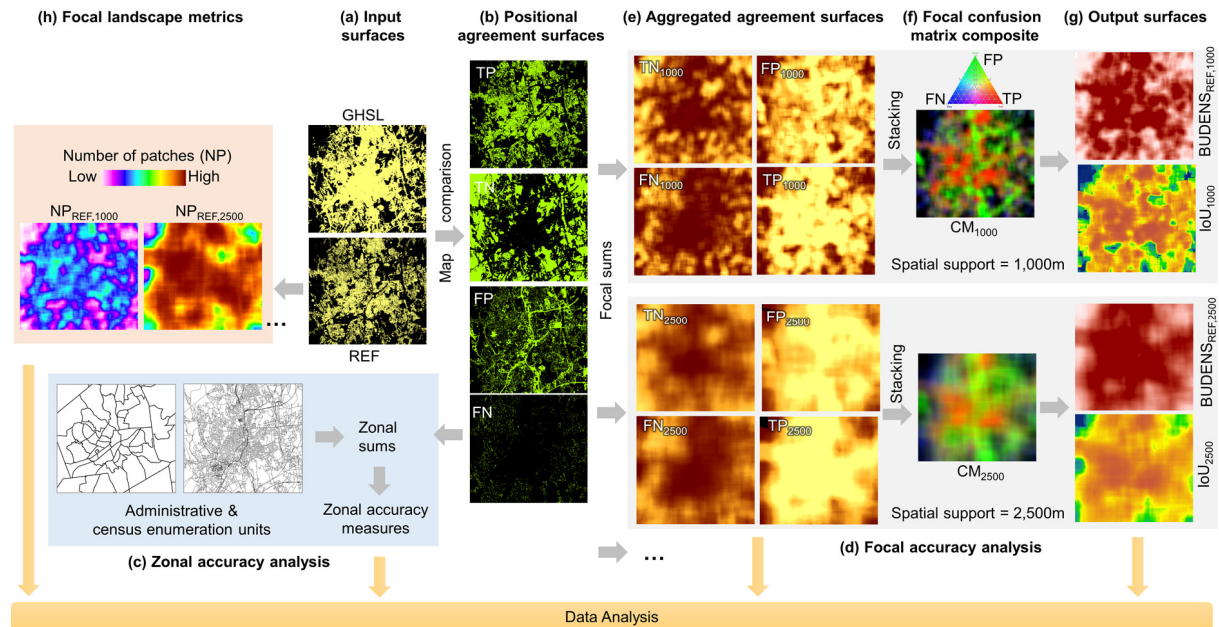

**Figure 2. Workflow of the conducted data processing steps: (a) Binary input surfaces indicating the presence and absence of built-up areas, (b) agreement surfaces for each agreement type (TP, TN, FP, FN) obtained by cell-by-cell map comparison, (c) zonal accuracy measure creation, (d) analytical steps for the focal accuracy analysis, shown for two of four levels of assessment support, (e) aggregated agreement surfaces, (f) resulting focal confusion matrix composites, (g) exemplary output surfaces generated from the confusion matrix composite, and (h) exemplary focal landscape metrics (i.e., number of patches, NP) derived for two levels of assessment support. Surfaces are shown for Worcester, Massachusetts.**

### 2.2.5    Assessing the effects of positional uncertainty in reference and GHSL data

The Landsat-based multispectral data used as input for the GHS-BUILT data has an approximate positional accuracy of 12-23m (Zanter 2017). In addition to that, the geodetic datum transformation applied when reprojecting the data into the target reference system can be expected to ingest additional positional uncertainty in the range of few meters. The building footprint data obtained from cadastral data sources, used to generate the reference surfaces are expected to have high positional accuracy, but may be affected by a spatial tolerance of up to 12m (Craig and Wahl 2003). Thus, the thematic accuracy estimates obtained from the gridded surfaces at the original resolution of 30m may be biased by misalignments due to the positional uncertainty in the underlying datasets (Congalton 2007). In order to mitigate this effect, we carried out some of our analyses based on 3x3 pixel blocks as assessment unit (e.g., Gu & Congalton 2020, Gu & Congalton 2021, Marconcini et al. 2020a) and analyzed how these results differ from the analyses carried out using individual grid cells for map comparison (Sections 3.5 and 3.6).

### 2.2.6    Analytical framework

Based on the generated data structures and surfaces (Figure 2), we carried out the subsequent analyses. Spatial data processing was done using Python 3.6, ESRI ArcPy Python package (ESRI 2020) and Geospatial Data Abstraction Library (GDAL; GDAL/OGR contributors 2020). The analytical steps are as follows: We examined the relationships of various (thematic and quantity) agreement measures characterizing zonal accuracy (Section 2.2.1) and focal accuracy (Section 2.2.2) with population density and built-up density to identify possible associations that could be useful for localized accuracy estimation of the GHSL across the rural-urban continuum (Section 3.1). Moreover, we conducted a theoretical suitability assessment of commonly used accuracy measures for small sample sizes and extreme class imbalance, and their plausibility with respect to theoretical expectations (i.e., the assumed increase of accuracy from rural towards urban settings) (Section 3.1).

Based on this assessment, we identified a set of suitable agreement metrics, for which the remainder of this analysis was carried out. We analyzed the interactions between omission and commission errors across the rural-urban continuum (Section 3.2) and examined the relationships between localized accuracy estimates and structural characteristics of built-up areas (Section 3.3). We then analyzed the sensitivity of zonal and focal localized accuracy estimates to the assessment support (Section 3.4). Subsequently, we tested the robustness of our analyses to the effects of positional uncertainty in reference and GHSL data, by applying selected analytical steps based on agreement metrics derived from the aggregated 3x3 pixel blocks (Section 3.5).

Finally, we applied our framework and integrated focal accuracy surfaces derived for the different GHSL epochs (i.e., 1975, 1990, 2000 and 2014) in order to assess how the localized GHSL accuracy varies over time (Section 3.6).

## 3.    Results

### 3.1   Suitability of agreement measures for localized accuracy estimation across the rural-urban continuum

The visualization of the generated (focal) accuracy surfaces (Section 2.2.2) allows for a visual comparison of the measures under test. These surfaces are shown for the Greater Worcester area in Figure 3. As can be seen in Figures 3a) and b), built-up areas are well detected in densely developed areas of the urban core, whereas peri-urban settlements are less well detected in GHSL. This trend has been observed in previous work (Leyk et al. 2018, Uhl et al. 2018) and constitutes important domain knowledge for the evaluation of these accuracy surfaces. The quantity agreement measures (Figure 3n and o) confirm this trend, reporting overestimation in urban areas, and underestimation in rural areas (Liu et al. 2020). Among the tested thematic agreement measures, this trend of increasing accuracy from rural towards more urban settings is only visible for Precision, Recall, F-measure, and IoU, (Figure 3d,e,f,h,l). The IoU exhibits the visually highest similarity to the reference built-up density surface (Figure 3c). Among these four accuracy measures, Recall shows the least variation within the urban core, indicating consistently low levels of omission errors for most parts of the urban core. A reverse trend can be observed for PCC due to class imbalance caused by the dominant not built-up class in rural settings (Figure 3i), whereas G-mean and Kappa (Figure 3j and k, respectively) report low levels of agreement in the densely built-up urban core, caused by the absence of negative instances (i.e., non-built-up grid cells) and consequently, low proportions of "true negatives" (cf. Eqs. 7 and 11). NMI (Figure 3m) as a conservative agreement measure exhibits low values, low levels of spatial variation, and is not defined in the urban core, where dense built-up grid cells in the reference data provoke the absence of false negatives. Similar effects are observed for the adjusted F-measure (Figure 3g) and MCC (Figure 3l). These observations indicate the need for agreement measures to characterize localized accuracy in more meaningful ways by being robust to class imbalance, insensitive to low values found in elements of the CM, and mathematically defined such that CM elements of value zero can be handled.

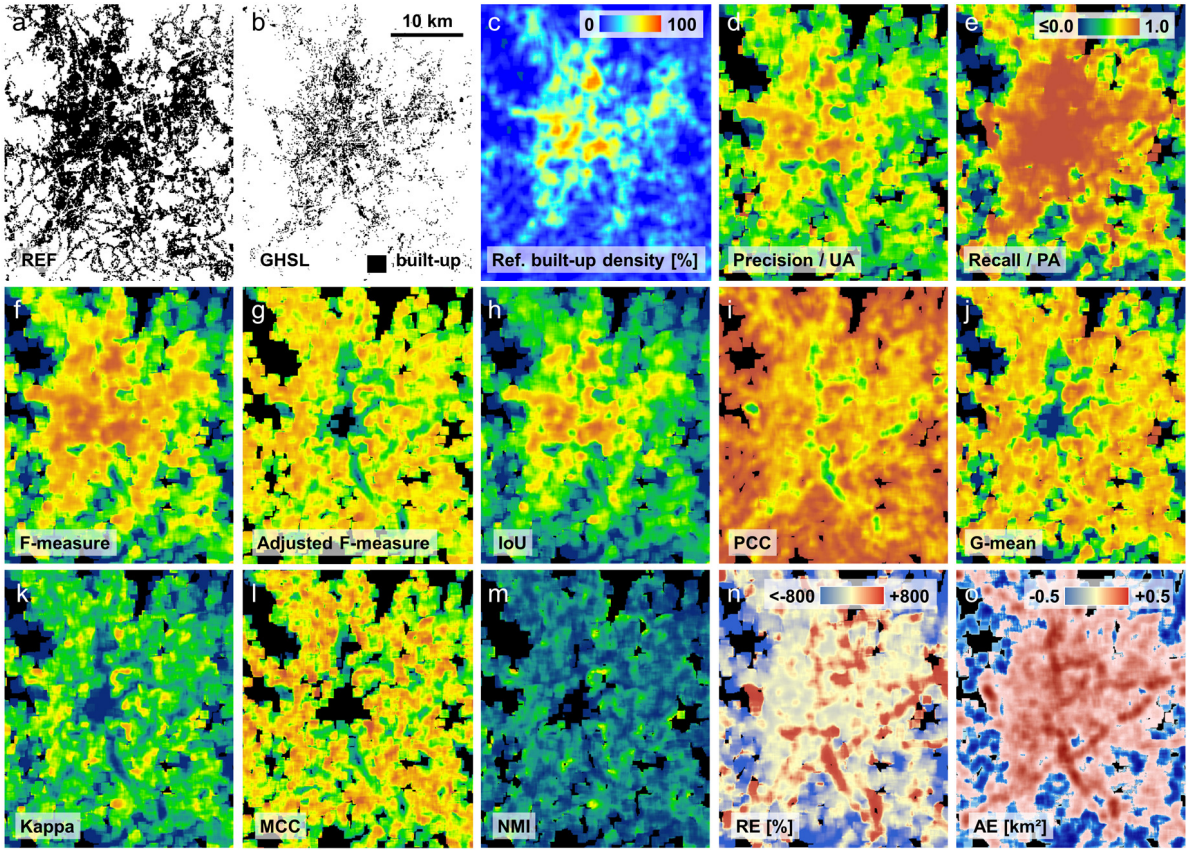

**Figure 3. Input data, derived focal density and accuracy surfaces for the agreement measures used herein, computed at spatial resolution of 30m, using an assessment support (i.e., focal window) of 1x1km: (a) Reference built-up labels, (b) GHSL 2014 built-up grid cells, (c) localized built-up density derived from the reference data, and surfaces of (d) Precision, (e) Recall, (f) F-measure, (g) Adjusted F-measure, (h) Intersection-over-union, (i) Percentage correctly classified, (j) G-mean, (k) Kappa, (l) Matthews correlation coefficient, (m) Normalized mutual information, and focal quantity agreement surfaces (n) relative error, and (o) absolute error.**

Table 1 shows the numerical robustness of the examined agreement measures for systematic combinations of empty confusion matrix elements. It is notable that both F-measure and G-mean are not defined in cases when the assessment support of a confusion matrix does not contain true positive instances. The desired output in such cases would be a value of 0. Additionally, the G-mean is not defined when false positive instances are absent. This is critical since both scenarios are common in sparsely built-up, and highly built-up areas, respectively. Also notable is that NMI is not defined if any element of the confusion matrix is zero. IoU yields valid outputs for 13 out of 14 cases, representing a promising, robust measure for such agreement assessments. While this experiment considers strict mathematical definitions, some of these non-definition problems can be mitigated by identifying the problematic cases and setting the measures to 0 (Chicco and Jurman 2020).

**Table 1. Numerical robustness table of the agreement measures used in this study. Each line represents a unique combination of presence / absence of the four agreement categories. The four blocks show different combinations of presence / absence of positive instances (i.e., TP and TN). Check marks indicate valid numerical values, hyphens denominate undefined instances. Values of 0.0 and 1.0 (and -1.0 for MCC) are indicated explicitly.**

| TP | TN | FP | FN | UA (Precision) | PA (Recall) | F-measure | Adj. F-measure | IoU | PCC | G-mean | Kappa | MCC | NMI | RE | AE |
|----|----|----|----|----|----|----|----|----|----|----|----|----|----|----|----|
| 0 | 0 | 0 | ✓ | - | 0.0 | - | - | 0.0 | 0.0 | - | 0.0 | - | - | ✓ | ✓ |
| 0 | 0 | ✓ | 0 | 0.0 | - | - | - | 0.0 | 0.0 | - | - | - | - | ✓ | ✓ |
| 0 | 0 | ✓ | ✓ | 0.0 | 0.0 | - | - | 0.0 | 0.0 | 0.0 | - | -1.0 | - | ✓ | ✓ |
| 0 | ✓ | 0 | 0 | - | - | - | - | - | 1.0 | - | 1.0 | - | - | - | 0.0 |
| 0 | ✓ | 0 | ✓ | - | 0.0 | - | - | 0.0 | ✓ | 0.0 | ✓ | - | - | ✓ | ✓ |
| 0 | ✓ | ✓ | 0 | 0.0 | - | - | - | 0.0 | ✓ | - | ✓ | - | - | - | ✓ |
| 0 | ✓ | ✓ | ✓ | 0.0 | 0.0 | - | - | 0.0 | ✓ | 0.0 | ✓ | - | - | ✓ | ✓ |
| ✓ | 0 | 0 | 0 | 1.0 | 1.0 | 1.0 | - | 1.0 | 1.0 | - | - | - | - | 0.0 | 0.0 |
| ✓ | 0 | 0 | ✓ | 1.0 | ✓ | ✓ | - | ✓ | ✓ | - | 0.0 | - | - | ✓ | ✓ |
| ✓ | 0 | ✓ | 0 | ✓ | 1.0 | ✓ | - | ✓ | ✓ | 0.0 | ✓ | - | - | ✓ | ✓ |
| ✓ | 0 | ✓ | ✓ | ✓ | ✓ | ✓ | - | ✓ | ✓ | 0.0 | ✓ | - | - | ✓ | ✓ |
| ✓ | ✓ | 0 | 0 | 1.0 | 1.0 | 1.0 | 1.0 | 1.0 | 1.0 | 1.0 | 1.0 | 1.0 | - | 0.0 | 0.0 |
| ✓ | ✓ | 0 | ✓ | 1.0 | ✓ | ✓ | ✓ | ✓ | ✓ | ✓ | ✓ | ✓ | - | ✓ | ✓ |
| ✓ | ✓ | ✓ | 0 | ✓ | 1.0 | ✓ | ✓ | ✓ | ✓ | ✓ | ✓ | ✓ | - | ✓ | ✓ |
| ✓ | ✓ | ✓ | ✓ | ✓ | ✓ | ✓ | ✓ | ✓ | ✓ | ✓ | ✓ | ✓ | ✓ | ✓ | ✓ |

To examine the visually observed trends in Figure 3 in a quantitative manner, we calculated Pearson's correlation coefficients for census tracts, block groups, and blocks between the localized accuracy estimates calculated for each of the >150,000 census enumeration units and the built-up density measures for each enumeration unit, as well as enumeration unit size and census 2010 population density, both reflecting the fine-grained population distributions (Table 3). General expectations, i.e., an unbiased agreement measure yielding high values in urban, and low values in rural areas, seems to be confirmed for the F-measure, IoU, Precision and Recall, yielding correlation coefficients of up to 0.95 for IoU, when compared to built-up density, and up to 0.60 when compared to population density. This implies two things: First, several agreement measures such as NMI, PCC, G-mean, MCC, adjusted F-measure, and Kappa do not seem to produce plausible results when tested against theoretical expectations. Second, among the agreement measures yielding geographically plausible results (F-measure, IoU, PA, UA), the IoU exhibits strongest levels of association to built-up and population density.

The quantity agreement measures AE and RE do not show such strong correlations. However, since AE and RE are composed of both omission and commission errors, we calculated separate correlation coefficients for the overestimation (OE) and the underestimation (UE) components for AE (Eqs. 17 and 18, respectively), as well as for the relative error RE (Eq. 16) (i.e., $UE_{REL}$, $OE_{REL}$). As shown in Table 2, as a result of this differentiation, correlation coefficients, in particular for the underestimation components, significantly increase as compared to AE and RE, respectively. These trends indicate that the degree of built-up quantity underestimation is negatively correlated with built-up density (i.e., GHSL is likely to underestimate built-up quantity in low built-up density areas, e.g., by omitting scattered rural settlements). In contrast, overestimation exhibits lower levels of correlation, and thus, seems to occur more independently from built-up density, e.g., due to roads and impervious surfaces misclassified as built-up land, which may occur in rural and urban regions.

**Table 2. Pearson's correlation coefficients of zonal accuracy measures and built-up densities, population density, and enumeration unit size, for census tracts, cencus block groups (CBGs), and census blocks.**

| Variable | Reference unit | Precision (UA) | Recall (PA) | F-measure | Adj. F-measure | IoU | PCC | G-mean | Kappa | MCC | NMI | AE | RE | $AE_{OE}$ | $AE_{UE}$ | $RE_{OE}$ | $RE_{UE}$ |
|---|---|---|---|---|---|---|---|---|---|---|---|---|---|---|---|---|---|
| Enumera-tion unit size | Tract | -0.432 | -0.615 | -0.573 | 0.082 | -0.510 | 0.390 | 0.054 | 0.011 | -0.115 | -0.077 | -0.231 | 0.044 | 0.552 | 0.637 | 0.134 | 0.487 |
| | CBG | -0.436 | -0.588 | -0.563 | 0.039 | -0.492 | 0.307 | 0.086 | 0.041 | -0.087 | -0.057 | -0.205 | 0.056 | 0.430 | 0.623 | 0.147 | 0.506 |
| | Blocks | -0.124 | -0.310 | -0.328 | 0.028 | -0.193 | 0.146 | 0.096 | 0.136 | 0.077 | 0.101 | -0.056 | -0.028 | 0.430 | 0.532 | 0.042 | 0.139 |
| Refe-rence built-up density | Tract | 0.893 | 0.789 | 0.913 | -0.443 | 0.945 | -0.084 | -0.576 | -0.371 | -0.188 | -0.227 | -0.090 | -0.071 | -0.495 | -0.470 | -0.091 | -0.475 |
| | CBG | 0.918 | 0.772 | 0.922 | -0.439 | 0.950 | 0.115 | -0.663 | -0.508 | -0.286 | -0.285 | -0.120 | -0.064 | -0.526 | -0.498 | -0.083 | -0.441 |
| | Blocks | 0.927 | 0.661 | 0.900 | 0.015 | 0.938 | 0.232 | -0.389 | -0.119 | -0.209 | -0.258 | -0.035 | -0.173 | -0.175 | -0.290 | -0.326 | -0.499 |
| GHSL built-up density | Tract | 0.731 | 0.897 | 0.855 | -0.565 | 0.863 | -0.362 | -0.583 | -0.431 | -0.223 | -0.212 | 0.041 | -0.057 | -0.416 | -0.526 | -0.083 | -0.571 |
| | CBG | 0.755 | 0.899 | 0.875 | -0.576 | 0.874 | -0.159 | -0.647 | -0.531 | -0.281 | -0.249 | 0.047 | -0.047 | -0.420 | -0.546 | -0.073 | -0.564 |
| | Blocks | 0.422 | 0.910 | 0.791 | -0.369 | 0.642 | -0.363 | -0.416 | -0.288 | -0.152 | -0.196 | 0.124 | 0.078 | -0.199 | -0.287 | -0.109 | -0.734 |
| Popula-tion density | Tract | 0.541 | 0.518 | 0.568 | -0.584 | 0.607 | 0.013 | -0.691 | -0.523 | -0.348 | -0.290 | -0.137 | -0.030 | -0.377 | -0.433 | -0.037 | -0.376 |
| | CBG | 0.513 | 0.476 | 0.535 | -0.500 | 0.570 | 0.142 | -0.662 | -0.520 | -0.323 | -0.234 | -0.122 | -0.025 | -0.345 | -0.439 | -0.032 | -0.394 |
| | Blocks | 0.314 | 0.309 | 0.390 | -0.034 | 0.360 | 0.133 | -0.267 | -0.101 | -0.105 | -0.129 | -0.026 | -0.051 | -0.086 | -0.192 | -0.095 | -0.349 |

The census-based, zonal accuracy estimates underlying the correlation analysis reported in Table 2 may suffer from a sampling bias, since census enumeration units are designed in a way that each unit contains a minimum population count. Assuming that the built-up area reflects the population counts to some degree, the existence of census enumeration units containing very few or no built-up instances in the test data is unlikely. Therefore, we cross-compared the relationships between built-up density and localized accuracy estimates based on zonal and focal constraining geometries (Figure 4). While illustrating the previously discussed variety in the trajectories of different agreement measures across the rural-urban continuum, these scatterplots exhibit high degrees of similarity between zonal (Figure 4a,b) and focal (Figure 4c,d) accuracy estimates. This indicates that, despite the above-mentioned sampling bias in census-based zonal accuracy estimates, the relationships to built-up density are of generalizable nature. More importantly, the trajectories of the respective agreement measures across the rural-urban continuum using reference built-up density (Figure 4a,c) and GHSL-based built-up density (Figure 4b,d) are highly similar, in particular when zonal geometries are used to define the assessment support. Comparing the F-measure and IoU, which exhibit strongest levels of correlation overall, the shape of the point clouds indicates a steeper slope of the F-measure, indicating higher levels of conservativeness of the IoU in low-density regions.

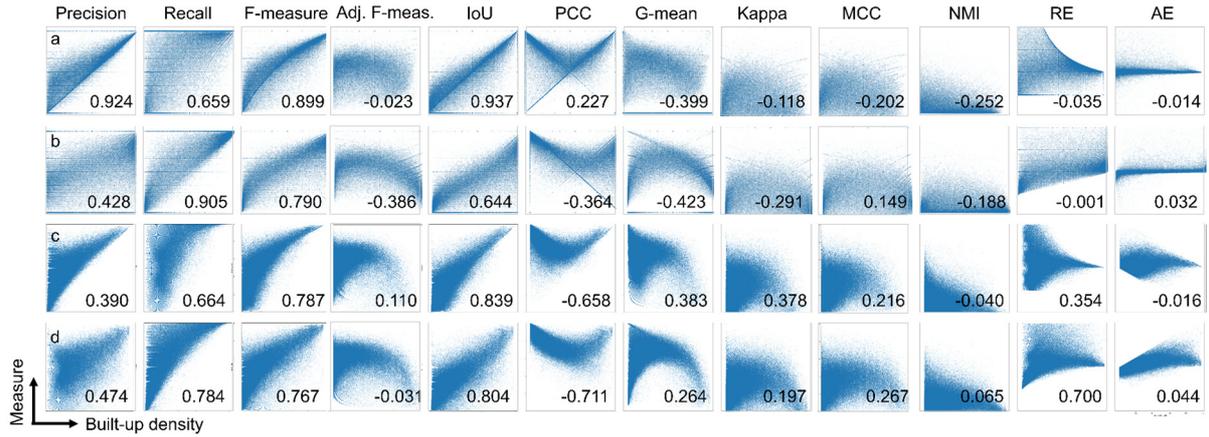

**Figure 4. Scatterplots of localized accuracy estimates (y-axis) against built-up density (x-axis): (a) zonal support, using reference built-up density, (b) zonal support, using GHSL-based built-up density, (c) focal support, using reference built-up density, (d) focal support, using GHSL-based built-up density. Underlying focal accuracy surfaces have a spatial resolution of 30m and are based on an assessment support of 1x1km. Zonal accuracy measures are shown for census tracts, block groups and blocks together. Also shown are Pearson's correlation coefficients for each scatterplot.**

Based on results shown in Figures 3 and 4, and Tables 1 and 2, we consider IoU the most suitable agreement measure for estimating the local accuracy variations of built-up land layers such as the GHSL, yielding geographically plausible and robust results, exhibiting strong associations with built-up density and population density, and thus, showing potentially high degrees of predictability.

### 3.2 Interactions of omission and commission errors across the rural-urban continuum

As the different patterns of precision and recall distributions against built-up density in Figure 4 suggest, omission and commission errors appear to follow different trajectories across the rural-urban continuum. In order to test this, we used our 1,000,000 sample locations drawn from the confusion matrix composites (cf. Figure 2f, Section 2.2.5). For each of these sample locations, we calculated the $F_\beta$ measure for a range of $\beta = 0.5$ to $\beta = 2.0$, in increments of 0.1. This allowed us to assess the variations of the $F_\beta$ measure on a continuous scale between the $F_{0.5}$ measure (favoring precision over recall) and the $F_{2.0}$ measure (favoring recall over precision), within quintile-based strata of reference built-up density (Figure 5a). The median $F_\beta$ trends per density stratum reveal interesting details: In the low-density stratum, both extremely low precision and recall values seem to occur, resulting in a symmetric, slightly U-shaped curve of median $F_\beta$ across the $\beta$ range. This is likely to be a superposed effect of highly precise built-up grid cells in GHSL, suffering from high omission errors, and a low-precision component induced by falsely labelled road grid cells. The effect of this low-precision component disappears in density stratum 2, where the median $F_\beta$ trend indicates low recall but high levels of precision. In the medium density stratum 3, precision and recall appear to be equilibrated. The trend is inverted in the high-density strata 4 and 5, where the $F_\beta$ measure decreases with higher weight to precision, and increases if more weight is given to recall, reflecting high levels of commission errors (e.g., roads, impervious surfaces) and low levels of built-up area omission in the GHSL. The supplementary movie illustrates the effect of $\beta$ on the relationship of the $F_\beta$ measure and built-up density.

Which $F_\beta$ measure best reflects the rural-urban gradient? We analyzed the correlations of the $F_\beta$ measure for a range of $\beta \in [0.1, 2.0]$, for all four levels of assessment support (Figure 5b), indicating maxima of Pearson's correlation coefficient for $\beta$ between 0.75 and approximately 0.9, for all support levels, suggesting that an $F_\beta$ measure slightly favoring precision exhibits a stronger linear relationship across the rural-urban continuum than the unweighted F-measure. Here, it is worth noting that none of these correlation maxima exceeds the correlation between IoU and reference built-up density of 0.84 (Figure 4c).

This indicates that precision and recall follow different trajectories across the rural-urban continuum. Thus, we analyzed the relationship between precision and recall themselves (Figure 5c), and found a much steeper increase of recall compared to precision. Figure 5c shows most locations in rural areas (i.e., low built-up density) are found in the lower triangle of the plot (i.e., precision > recall), whereas in higher density regions, recall seems to be greater than precision (upper triangle). This asymmetric relationship between precision and recall is also reflected in the ternary plot shown in Figure 5d, which is based on the relative proportions of TP, FP, and FN at each sample location. Figure 5d also shows locations where precision equals recall, which we call "isometric" locations, which can be found for a wide range of TP, and throughout the rural-urban continuum, except in areas of high built-up density. These locations are particularly interesting, as the quantity error is 0, and thus, regardless the level of thematic disagreement, the GHSL provides correct estimates of total built-up area. Moreover, we identified locations where the relative difference between precision and recall does not exceed 10%, and 20%, respectively, and visualized the distribution of these "quasi-isometric" locations along the rural-urban continuum (Figure 5e). These kernel density functions indicate that quasi-isometric locations are mainly found in rural regions with built-up densities of 5-15%, and this peak is more nuanced as assessment support increases.

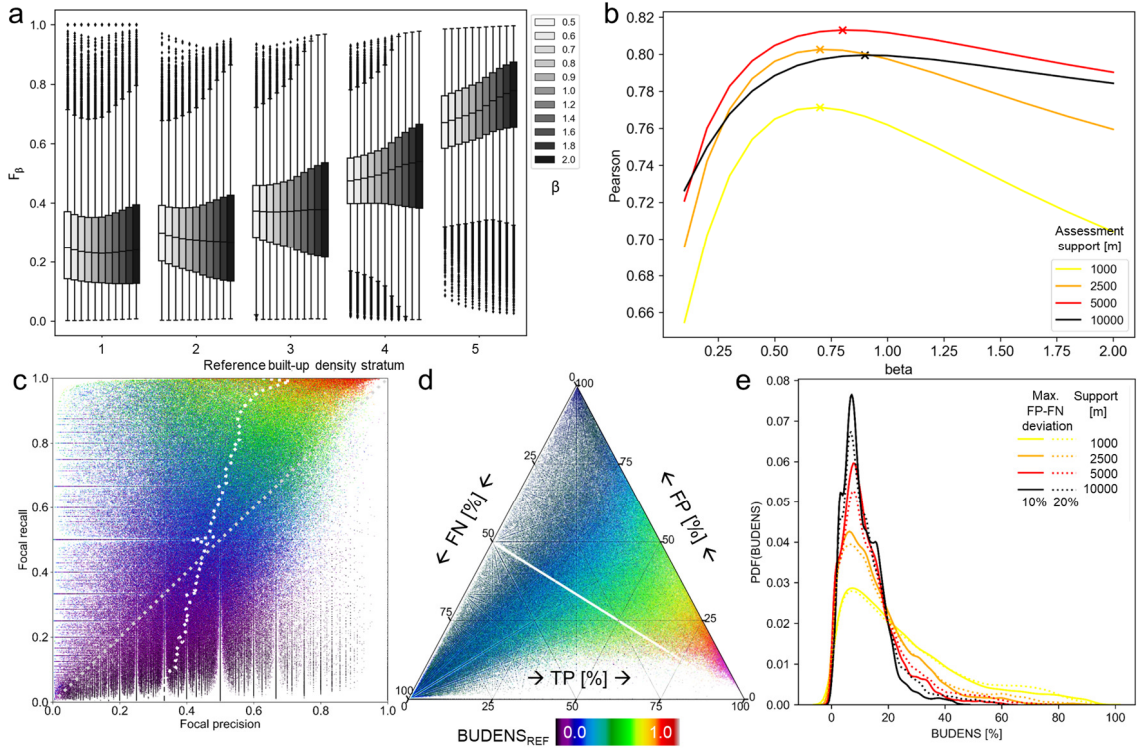

**Figure 5. Interactions of precision and recall across the rural-urban continuum: (a) Distributions of the $F_\beta$ measure for different values of β, within strata of reference built-up density, (b) Pearson's correlation coefficient between reference built-up density and $F_\beta$ for a range of β values, (c) Localized precision-recall scatterplot, color-coded by reference built-up density, white dashed line represents the average precision per recall percentile. (d) True positive, false positive, and false negative ternary plot, color-coded by reference built-up density; including locations where precision equals recall in white and (e) reference built-up density distributions (i.e., probability density functions) within bands of precision-recall similarity (i.e., 10% and 20% maximum deviation between precision and recall). Ternary plot in (d) created with python-ternary (Harper et al. 2015).**

### 3.3 Interactions between localized accuracy estimates and density-/structure-related characteristics of built-up surfaces

While built-up density represents a commonly used and computationally inexpensive proxy variable to characterize the rural-urban continuum, structural measures describing the shape and spatial segregation of built-up areas may relate differently to localized accuracy estimates. To explore this, we explicitly analyzed two landscape metrics and their relationship to built-up density, the IoU as a thematic accuracy measure, as well as the absolute error (AE) as a measure of quantity agreement. These landscape metrics include the number of contiguous built-up area patches (NP) and the largest patch index (LPI, reflecting area proportion of the largest built-up patch), computed at sample locations within focal windows of 1×1 km (see Section 2.2.5). Comparison of these structural measures with the focal IoU surface (cf. Figure 3h) and the corresponding built-up density surface allow for visualizing the rural-urban continuum in two-dimensional spaces (Figure 6): Figure 6a shows the interactions of built-up density, NP and the IoU, indicating high thematic accuracy where built-up density is high and number of patches is low (i.e., dense, contiguous patches of built-up land, such as urban cores), decreasing towards peri-urban areas (i.e., moderate built-up density, high levels of segregation), and rural areas (low built-up density, and few, scattered settlement patches). The visualization of quantity agreement (AE, Figure 6b) reveals that the underestimation of built-up area (i.e., negative AE) mainly occurs in areas characterized by low and moderate built-up density, but relatively independently from the level of spatial segregation of built-up areas. However, the shapes of the point clouds in Figure 6a,b illustrate the ambiguous nature of the NP metric to characterize the rural-urban continuum, as the same values of NP can be found in both, low and high-density regions. Combining built-up density and LPI shows a different picture: Thematic accuracy of the GHSL is mainly driven by built-up density, and occur for both, large and small contiguous patches of built-up land (Figure 6c). Underestimation of built-up area (i.e., quantity disagreement measured by negative AE) occurs mostly in areas of low built-up density characterized by small patches of built-up land (Figure 6d).

Lastly, we assessed the interactions between thematic agreement (IoU) and quantity agreement (AE) across density and structure of the built-up areas (Figure 6e-h). As expected, we observe a general trend of decreasing AE with increasing IoU, across all strata of LPI. In regions of low LPI (i.e., small, scattered patches of built-up land, Figure 6e), we mainly observe built-up land overestimation (i.e., AE>0), possibly due to highways and roads misclassified as built-up land in the GHSL. In other words, low IoU values in these regions are driven by high proportions of false positives. Conversely, in regions where large, contiguous patches of built-up land dominate (i.e., high LPI, Figure 6h), we observe higher levels of overestimation despite moderate or high IoU. Highest IoU values occur in the high LPI stratum, which is in agreement with previous work (Klotz et al. 2016, Mück et al. 2017).

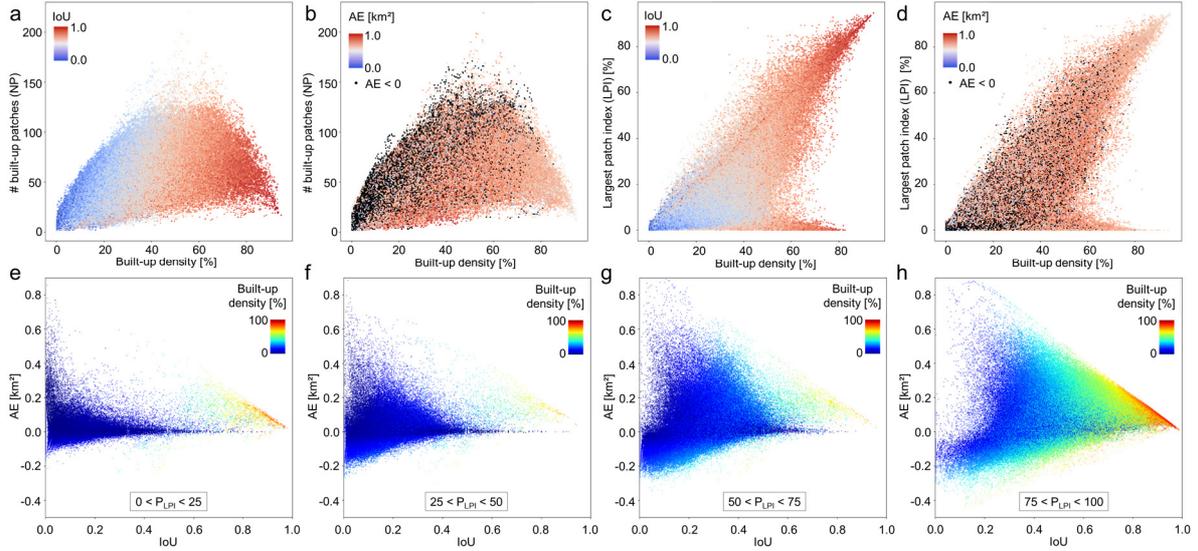

**Figure 6. Interactions between thematic accuracy, quantity agreement, and density-structure characteristics of built-up land: (a) IoU and (b) AE color-coded in a bi-dimensional space of built-up density and number of patches, (c) IoU and (d) AE color-coded in a bi-dimensional space of built-up density and LPI, (e – h): built-up density color-coded in a bi-dimensional space of IoU and AE, stratified based on percentiles of LPI ($P_{LPI}$). All built-up density and structural variables are derived from the reference data, localized measures are based on a focal window of 1x1km.**

### 3.4  Assessing sensitivity of localized accuracy estimates to assessment support

Up to this point, our analysis was based on localized accuracy estimates derived from fixed levels of assessment support, without taking into account potential sensitivity of these estimates to assessment support. In this section, we aim to identify such sensitivities. First, we visualize localized accuracy estimates derived from the zonal geometries (see Section 2.2.1) in geographic space. Mapping the IoU at different levels of assessment support illustrates the inherent spatial variability across different geographical extents (Figure 7). Whereas IoU at the state level (Figure 7a) has a similar magnitude as the majority of counties (Figure 7b), it decreases in most entities of the subsequent finer scales (Figure 7c-f), especially in rural settings. In highly urban regions (e.g., Greater Boston), IoU tends to increase from state to census tract level but yields highly dispersed values when using units of finer granularity. Thus, IoU generated from the state level-confusion matrix underestimates thematic agreement in urban settings and overestimates in rural areas. The low IoU in rural settings is likely due to a high number of false positives caused by road features detected as built-up land in GHSL (cf. Figure 6e), alongside with high levels of omission errors caused by the difficulty in detecting dispersed small settlements in GHSL.

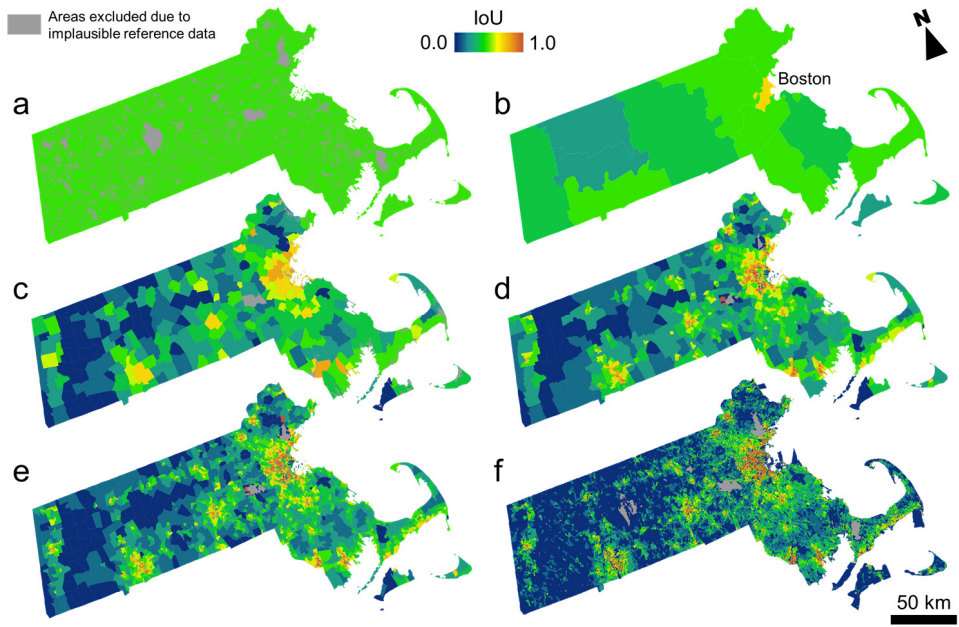

**Figure 7. IoU for different levels of assessment support derived from constraining zonal geometries: (a) State, (b) counties, (c) townships, (d) census tracts, (e) census block groups, and (f) census blocks. Gray areas in (a) are excluded from the analysis due to implausible reference data.**

Whereas such map sequences across levels of assessment support illustrate the spatial variability of the accuracy estimates and their support dependency, it is difficult to detect and visualize cross-support effects. We generated cross-support trajectory plots for thematic agreement (IoU, Figure 8a) and quantity agreement (AE, Figure 8b) and, for cross-comparison, for Kappa and PCC (Figures 8c and d, respectively), for all 157,508 census blocks in Massachusetts, and observe the following:

**Support sensitivity.** Among the shown thematic accuracy measures, IoU exhibits the widest range of magnitudes, and shows lowest degrees of sensitivity (i.e., high stability) across all support levels from township to block group level. This implies that the proportion of misclassified instances stays stable across these assessment support levels. Kappa exhibits a considerable amount of trajectories dropping to very low values from township to block group level, and converging to extreme values (i.e., 0 and 1) at the block level. This indicates high levels of support sensitivity and confirms common criticisms to Kappa, such as its sensitivity to marginal probabilities (Gwet 2002), or its non-suitability for accuracy assessments of land cover data (Pontius and Millones 2011, Foody 2020).

**Sample size sensitivity.** All accuracy measures under test show high levels of diffusion at the lowest level of assessment support, the block level. This indicates high degrees of sensitivity to small sample sizes, taking into account that the median size of census blocks in Massachusetts is 16,175 sqm, corresponding to a sample size of 18.0 grid cells of 30x30m to establish the confusion matrix (2.4 grid cells for the 25th percentile, and 42.9 grid cells for the 75th percentile, respectively). Moreover, the accuracy values tend to take extreme values (i.e., 0.0, 1.0) due to critically low sample sizes and a lack in robustness of the accuracy measures when using low sample sizes.

**Trend.** PCC and Kappa exhibit decreasing trends towards the block level. While a decreasing trend for PCC can be explained with an increase in class balance, Kappa is showing a nearly linearly decreasing trend towards the block group level, i.e., with decreasing sample size. Such a trend, alongside the previously observed diffuse behavior for small sample sizes are in line with earlier work examining critical sample sizes and sample size dependency for inter-rater agreement measures such as the Kappa index (e.g., Sim and Wright 2005, Bujang and Baharum 2017). Regarding the average trajectories for rural and urban census blocks, the IoU shows the geographically most plausible picture, i.e., higher levels of accuracy in census blocks of high built-up density, as observed in Section 3.1.

**Conservativeness.** Whereas Kappa exhibits lower magnitudes across the examined support levels down to the block group level and thus, characterizes accuracy in a rather conservative way, PCC tends to yield high values that decrease steadily down to the block group level, confirming the well-known issue of PCC to yield inflated values, in particular when the evaluated classes are imbalanced (Rosenfeld and Melley 1980, Shao et al. 2019, Stehman and Wickham 2020). Moderate PCC values (here, approx. 0.6 – 0.8) yielded for township, tract and block group level could indicate a more balanced class distribution of built-up and not built-up classes.

The AE as a quantity agreement measure exhibits a distinct picture: As expected, AE as an absolute measure decreases with decreasing assessment support (i.e., towards block level), consistently for most census blocks. Several trajectories switch sign, indicating that the change in assessment support can cause a switch from under- to overestimation or vice-versa, likely due to heterogeneous or disproportionate levels of built-up density within those zones.

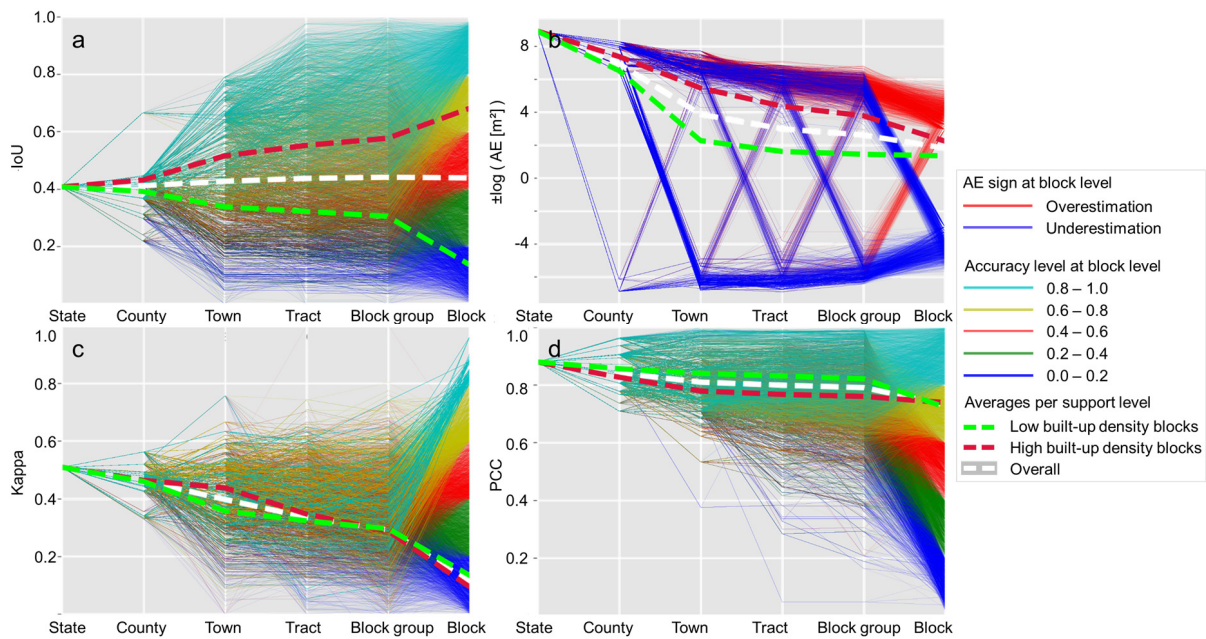

**Figure 8.** Cross-support trajectories of (a) IoU, (b) AE, (c) Kappa, and (d) PCC for all 157,508 census blocks in Massachusetts. Trajectories are color-coded by the agreement level at census block scale. Dashed lines represent average trajectories for census blocks overall and in strata of low and high built-up density, using the 75th percentile as a separation threshold. For readability purposes, AE values are log-transformed while preserving the original sign (Webber 2012).

The census unit boundaries used to generate these zonal accuracy estimates typically align with human-made features (e.g., neighborhoods, major roads) and, less frequently with rivers, and thus, contain low levels of within-unit land cover variability, especially in urban and peri-urban areas. This circumstance may affect cross-support trajectories and introduce certain bias. Hence, we performed a similar analysis based on focal accuracy estimates across our set of accuracy surfaces for different levels of assessment support (see Section 2.2.2), allowing for cross-support trajectory analysis independently of externally imposed zoning boundaries.

More specifically, we extracted IoU, AE, Kappa, and PCC trajectories for our stratified random sample of 1,000,000 locations in Massachusetts across all levels of assessment support (Figure 9). These boxplots show the distribution of the accuracy measures across the rural-urban continuum, separately for each level of assessment support. The IoU trajectories (Figure 9a) confirm the trends observed in zonal accuracy trajectories across support levels (cf. Figure 8a): i) given any level of assessment support, accuracy increases with increasing built-up density; ii) IoU exhibits low levels of variance to the chosen support level in dense, urban areas (i.e., the increase of distribution medians with increasing support is least pronounced in the high-density stratum), and IoU dispersion (i.e., inter-quartile ranges) appears quite constant across support levels, and even across the rural-urban continuum. This implies that using the IoU to characterize localized thematic accuracy of built-up land data is largely invariant to the chosen level of assessment support in urban areas (unless very low assessment support levels are used, such as census blocks in highly populated urban areas; cf. Figure 8a), but may be sensitive to the level of support in rural regions.

The AE computed within focal windows of varying assessment support (Figure 9b) shows a distinct pattern. Average AE magnitudes and dispersion increase with increasing assessment support, across all density strata. This is expected, since AE is an absolute measure. Median AE across support levels decreases in rural strata, and increases in the more urban strata. In the high-density stratum, there are a few locations of quantity underestimation (i.e., negative AE), increasing with support, shown as outliers. These results highlight that localized accuracy measures such as the IoU and AE need to be interpreted carefully when the underlying assessment support is not constant.

Kappa (Figure 9c) shows a moderately increasing trend across the rural-urban continuum for all support levels. This trend is less pronounced than the IoU (Figure 9a), likely due to the numerical problems of Kappa in highly urbanized areas (cf. Figure 3k), while sensitivity to assessment support is similar to the IoU. PCC (Figure 9d) exhibits an opposite trend (i.e., decreasing accuracy from rural to urban areas), which emphasizes the previously discussed unsuitability of PCC for localized accuracy estimation (cf. Figure 3i). Notably, PCC exhibits the lowest level of assessment support sensitivity among the four accuracy measures.

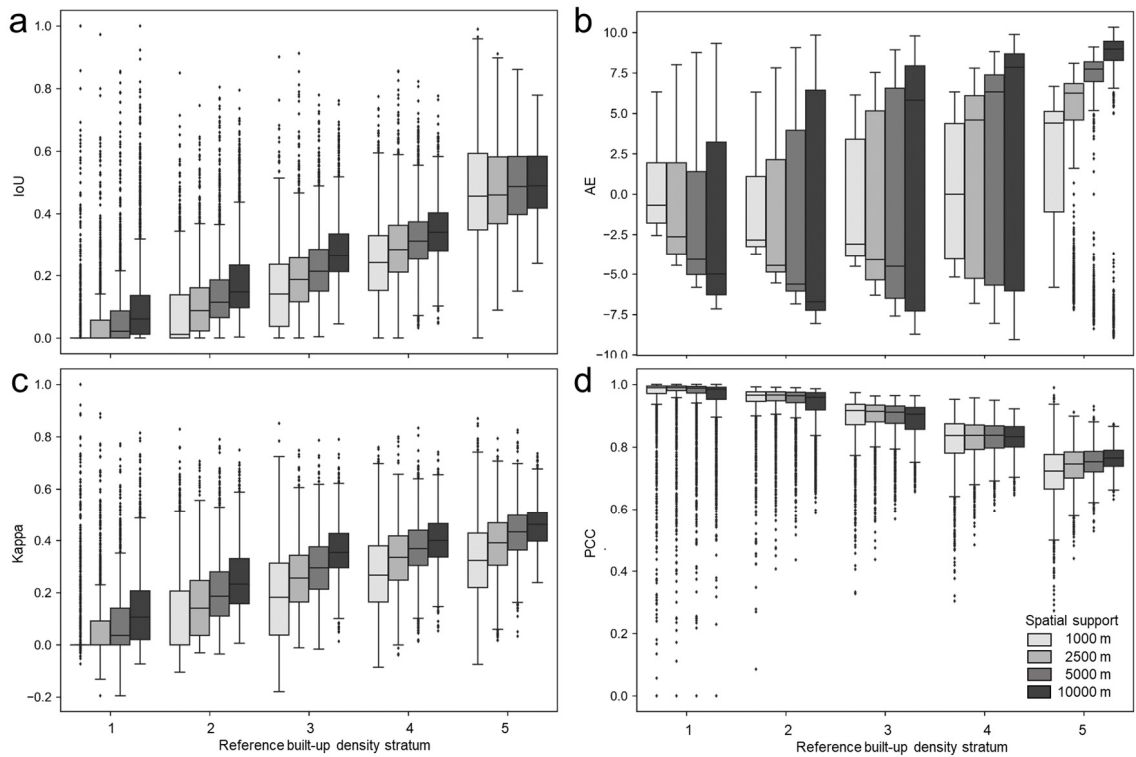

**Figure 9. Distributions of selected localized accuracy measures across strata of built-up density, computed within 1,000,000 focal windows for four levels of assessment support: (a) Intersection-over-union, (b) absolute error, (c) Kappa, and (d) PCC. For readability purposes, AE values are log-transformed while preserving the original sign (Webber 2012).**

While these distributions indicate considerable levels of sensitivity to assessment support, the correlations of these measures to built-up density appear to be stable across support levels for some measures (i.e., IoU, F-measure, precision, and recall), and increase with assessment support for the remaining measures, most notably for the AE and MCC (Figure 10a). This indicates that these measures may be suitable for localized accuracy characterization if assessment support / sample size of the underlying confusion matrices is large enough. For the accuracy measures exhibiting highest levels of correlation, correlation ranks are stable across the support levels, indicating high levels of generalizability of these relationships across spatial scales. Correlation trends with respect to the GHSL-derived built-up densities (Figure 10b) largely show similar trends for most measures, except the absolute error (AE) that shows a considerable increase in correlation. We also calculated these trends for the average of reference data and GHSL-derived built-up densities, and observe similar trends, ensuring that the data source for built-up density calculation does not affect our findings.

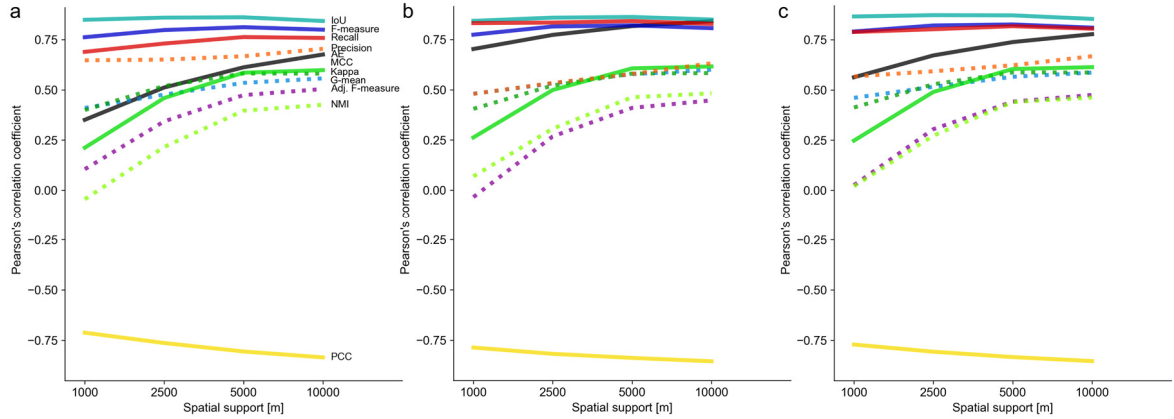

**Figure 10. Pearson's correlation coefficients for the analyzed accuracy measures to different versions of built-up density across levels of assessment support: (a) Built-up density derived from reference data, (b) derived from the GHSL, and (c) average of the former two.**

How does assessment support affect the relationships between accuracy measures? To investigate this, we visualized some of the previously discussed relationships at different levels of assessment support (Figure 11, see also Figure A1 for all support levels). For example, we observe an increasingly linear relationship between the IoU and AE, in particular in regions of higher built-up density, as assessment support increases from 2.5km (Figure 11a) to 10km (Figure 11b). This applies also to the relationship between precision and recall (Figure 11c,d) and the relative proportions of TP, FP, and FN (Figure 11e,f), reflected in a "bundling" effect. Correlation coefficients between these measures consistently increase with increasing support as well (Table 3). Thus, as assessment support increases, thematic and quantity agreement, as well as commission and omission errors, and agreement / disagreement measures become increasingly correlated to one another, and the fine nuances between different uncertainty types disappear with increasing assessment support. These results clearly demonstrate the need for localized accuracy estimates, revealing fine-grained uncertainty patterns that remain hidden if "global" or spatially over-generalized accuracy estimates are reported.

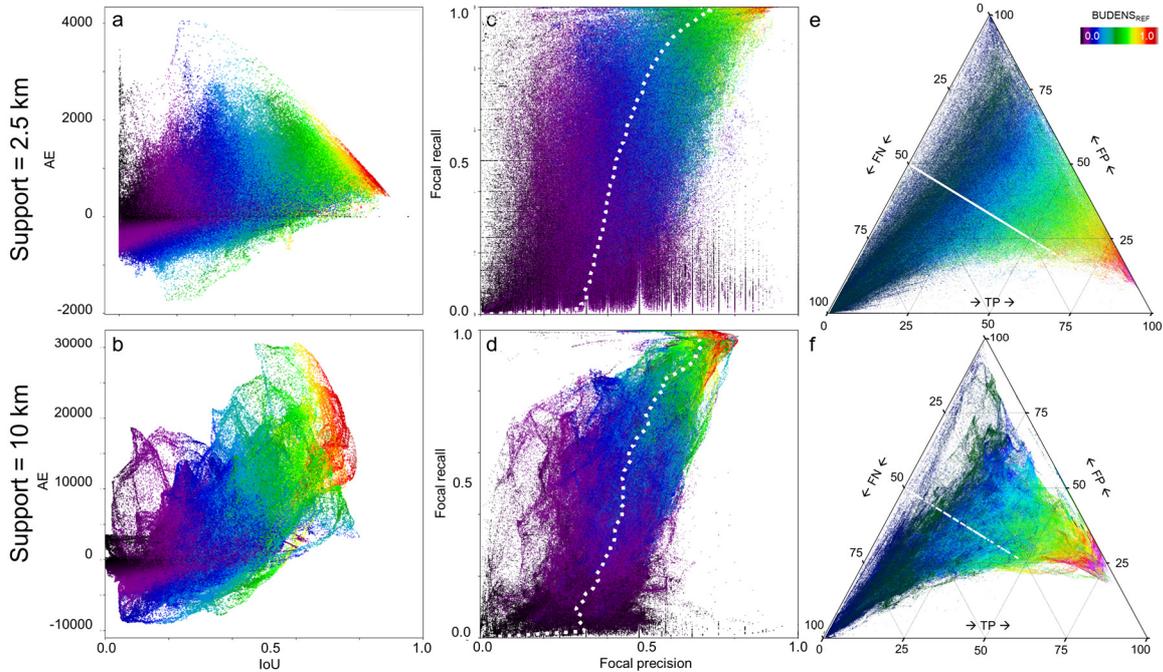

**Figure 11. Scatterplots of the relationships between selected localized accuracy measures and confusion matrix components, all color-coded by reference built-up density: (a,b) IoU versus AE, (c,d) Precision versus Recall, and (e,f) ternary plot of relative proportions of TP, FP, and FN. Top row: assessment support of 2.5km, bottom row: assessment support of 10km.**

**Table 3. Pearson's correlation coefficients between selected localized accuracy measures and confusion matrix components, for two levels of assessment support.**

| Measure 1 | Measure 2 | 1x1km | 10x10km |
|-----------|-----------|-------|---------|
| IoU | AE | 0.457 | 0.670 |
| Precision | Recall | 0.440 | 0.666 |
| FN | FP | 0.244 | 0.472 |
| TP | FP | 0.632 | 0.833 |
| TP | FN | 0.215 | 0.333 |

### 3.5 Analysis of assessment unit sensitivity

As described in Section 2.2.5, thematic accuracy estimates obtained from map comparison at the original resolution may be biased by the misalignment of gridded test and reference data, induced by the positional uncertainty of the underlying spatial data. This effect is expected to be mitigated by using 3x3 grid cell (i.e., 90m x 90m) blocks as assessment unit. Figure 12 a, b, and c show how IoU increases when using such aggregated units, particularly in sparsely built-up, peri-urban and rural areas (cf. reference built-up density surface in Figure 12d). This effect causes an average increase of IoU of about 0.25 in medium-density regions (Figure 12e), likely a superposed effect of increasing levels of accuracy, and a decreasing aggregation effect from rural to urban areas. However, in relative terms, this effect is most nuanced in sparsely built-up rural areas (Figure 12f). These trends are persistent over time when comparing the epochs 1975 and 2014 (Fig 12 e,f).

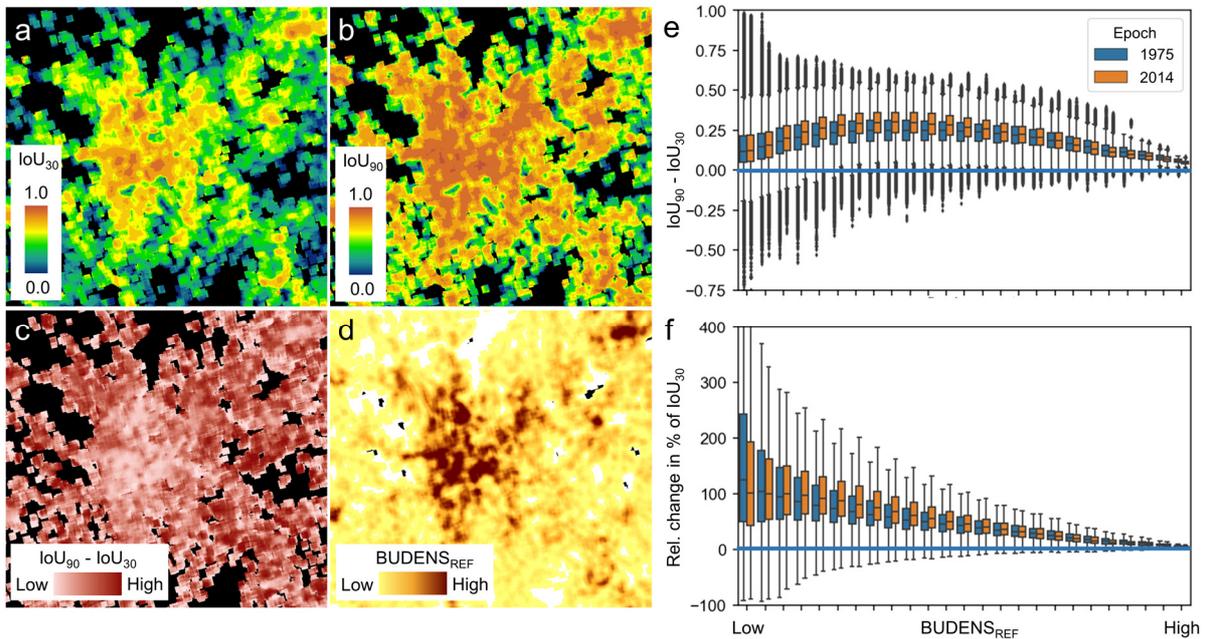

**Figure 12. Quantifying the bias in the focal thematic accuracy estimates introduced by positional uncertainty in the data. Focal IoU surface using (a) 30m individual grid cells and (b) 3x3 grid cell (i.e., 90m x 90m) blocks as assessment unit. Panel (c) shows the pixel-wise difference of IoU (90m blocks) and IoU (30m cells), and (d) reference built-up density surface for comparison, all shown for the city of Worcester, Massachussets). Boxplots show the trends of these differences across the rural-urban continuum, modelled by a percentile-based classification of the reference built-up density: (e) absolute IoU difference, and (f) difference in % of the IoU based on 30m cells as assessment unit, both shown for the GHSL epochs 1975 and 2014. The blue line in (e) and (f) indicates a difference of zero.**

The IoU obtained from map comparison at an assessment unit of 3x3 grid cell blocks is expected to be more realistic, as it is free from bias introduced by positional uncertainty. Since this effect is more pronounced in rural areas, the "true" trend of IoU across the rural-urban continuum (see Figure 4) is expected to be less steep, and the "true" correlation between IoU and built-up density is expected to be lower. This is confirmed by the scatterplots shown in Figure A2 and the correlation coefficients reported in Table 4. However, Table 4 shows that IoU exhibits higher levels of correlation to built-up density than the F-measure, and thus, indicates that our observations made in Section 3.1 are unaffected by potential bias due to positional uncertainty in the data. Moreover, these trends are highly persistent over time (Table 4).

**Table 4. Pearson's correlation coefficients of the IoU and F-measure with reference built-up density, for different assessment units and the epochs 1975 and 2014.**

| Assessment unit | Accuracy metric | Correlation w/ ref. built-up density 1975 | Correlation w/ ref. built-up density 2014 |
|---|---|---|---|
| 30x30m cells | F-measure | 0.761 | 0.759 |
| 3x3 cell blocks | F-measure | 0.665 | 0.675 |
| 30x30m cells | IoU | 0.812 | 0.810 |
| 3x3 cell blocks | IoU | 0.733 | 0.745 |

### 3.6 Assessing focal accuracy over time

The observations made in Section 3.5 regarding sensitivity to the assessment unit and accuracy trends across the rural-urban continuum appear to be highly persistent over time. But how does the local accuracy of the GHS-BUILT surfaces change over time, and how do such temporal trends play out across space and along the rural-urban continuum? To answer this question, we visualize IoU trends across the four epochs 1975, 1990, 2000, and 2014 for three strata based on reference built-up density in 2014. The thresholds for this stratification are adopted from a strategy used in Leyk et al. (2018), where two sets of thresholds were applied in order to ensure that the choice of thresholds does not affect the resulting trends. We observe mostly increasing thematic accuracy trends over time, for both assessment units. The decreasing trend of IoU over time in the low-density stratum (Figure 13a) indicates that opposite behaviour may occur, likely due to increased construction activity of scattered, rural settlements within our study area during the period 1975-2014, alongside with low sample sizes in the rural stratum.

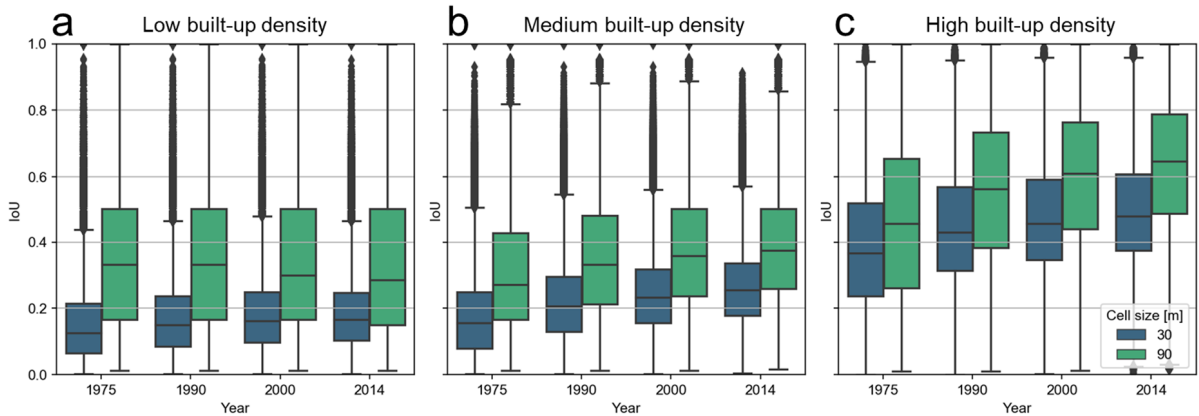

**Figure 13. Trends of IoU across the four GHSL epochs 1975 – 2014, within strata of reference built-up density, loosely related to (a) rural (0%-5% built-up density), (b) peri-urban (5%-20% built-up density), and (c) urban (>20% built-up density).**

These areas constitute around 36% of the land area within the rural stratum, and the average built-up density in these areas is low (2,5%), as shown in Table 5. These statistics represent a refined, more localized insight into GHSL accuracy trends over time, reported in Leyk et al. (2018), indicating that accuracy increased in most areas in both rural and urban regions, and accuracy increase tends to occur in the regions of higher built-up density within each stratum. We observe similar trends when using IoU based on 3x3 cell blocks, again confirming that these trends appear to be robust to changes in the assessment unit or underlying positional uncertainty.

**Table 5. Change statistics of localized IoU from 1975 to 2014, within strata of reference built-up density, for both single cells, and 3x3 cell blocks used as unit for the accuracy assessment.**

| Stratum | IoU temporal trend | Analytical unit = 30m cells | | | Analytical unit = 3x3 cell blocks | | |
|---|---|---|---|---|---|---|---|
| | | Area proportion [%] | Avg. ΔIoU | Avg. built-up density [%] | Area proportion [%] | Avg. ΔIoU | Avg. built-up density [%] |
| Low-density | Increasing | 63.14 | 0.08 | 3.17 | 50.86 | 0.09 | 3.39 |
| | Decreasing | 36.86 | -0.06 | 2.58 | 49.14 | -0.15 | 3.15 |
| Medium-density | Increasing | 85.02 | 0.11 | 11.97 | 70.93 | 0.16 | 13.86 |
| | Decreasing | 14.98 | -0.04 | 10.58 | 29.07 | -0.11 | 12.71 |
| High-density | Increasing | 95.11 | 0.12 | 37.51 | 89.49 | 0.19 | 53.13 |
| | Decreasing | 4.89 | -0.03 | 30.58 | 10.51 | -0.06 | 42.88 |

As mentioned above, we extracted these statistics for a second set of thresholds to establish rural, peri-urban and urban strata and generally observe similar trends, except in the rural stratum, which is likely an artefact of lower sample sizes (Figure A3, Table A2). In order to assess, how these accuracy trends play out across space, we visualized the focal confusion matrix composite (cf. Figure 2f) and the derived focal IoU surface for Greater Boston, both for the years 1975 and 2014, and for both assessment units (Figure 14a and b). These surfaces, and the difference surface shown in Figure 14c illustrate how accuracy increased notably in peri-urban regions around the city of Boston, and less so in dense urban areas.

This visualization, and the average built-up densities reported in Table 5 suggest that densification (i.e., change of reference built-up density over time) could be a driver for the increases in thematic accuracy. Based on the built-up density surfaces from the GHSL and the reference data extracted for 1975 and 2014 (Figure 14d) we generated surfaces of the densification per grid cell and observe considerable differences (Figure 14e). GHSL-based densification trends appear to overestimate the actual densification, in particular in peri-urban settings (Figure 14f), as a result of the higher omission errors in medium and low-density areas in the 1975 GHSL epoch.

Moreover, we observe that the reference densification surface appears very similar to the IoU difference surface (Figure 14g). The scatterplots in Figure 14h exhibit a relatively strong association between increasing built-up density and increased levels of thematic accuracy of the GHSL. Once again, this trend appears to be unaffected by positional uncertainty (Pearson = 0.46 for 30m-based IoU, and 0.44 for 90m-based IoU).

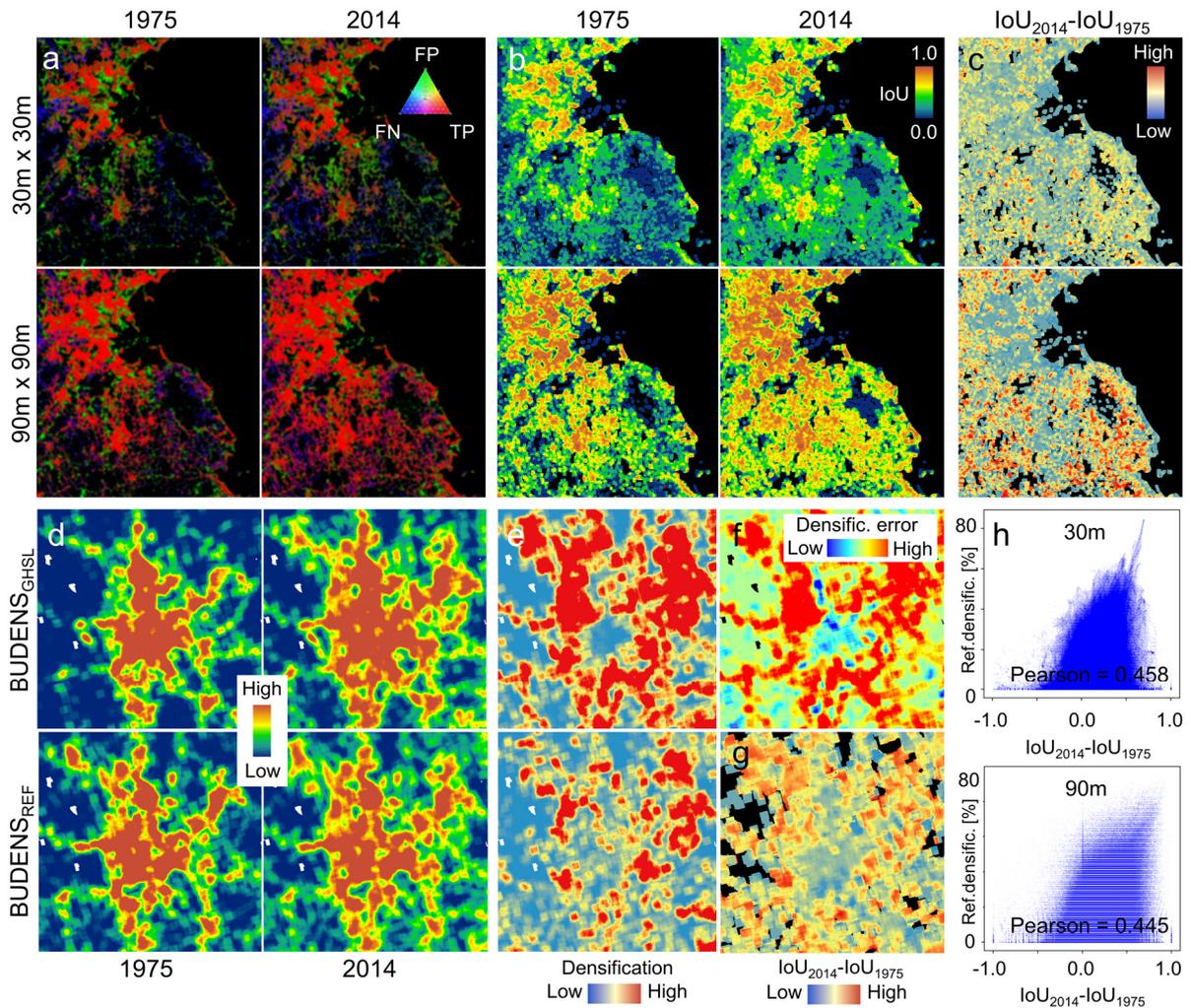

**Figure 14. Focal thematic and quantity agreement trends over time. (a)** Focal confusion matrix composite for 1km×1km assessment support, in 1975 and 2014, for both 30m grid cell and 90m blocks as assessment unit. **(b)** respective focal IoU surfaces, and **(c)** cell-wise IoU increase from 1975 to 2014. Panel **(d)** shows the focal built-up density surfaces in 1975 and 2014 derived from the GHSL and from the reference data, and **(e)** shows the corresponding densification surfaces. Panel **(f)** shows the absolute error of densification, and **(g)** shows the delta IoU for comparison. Panel **(h)** shows scatterplots and correlation coefficients of IoU increase over time and reference densification for both assessment units. Panels **(a) – (c)** show greater Boston, Massachusets, and panels **(d)-(g)** show the city of Worcester, Massachusetts. Scatterplots in **(h)** are based on all grid cells in Massachusetts.

## 4.  Discussion and conclusions

**Methodological contributions.** Herein, we presented a framework for the localized accuracy assessment of binary built-up surface datasets, which takes into account the peculiarities of such data products, i.e., shifting class imbalance across the rural-urban continuum. The proposed framework entails the creation of a set of focal density surfaces, counting the occurrences of grid cells per agreement category (e.g., TP, FP, FN, etc.) in a confusion matrix within focal windows of varying size (i.e., assessment support) (Figure 2). From a technical point of view, the proposed computational framework allows storing the elements of localized confusion matrices in data cubes, and enables efficient, exhaustive and spatially explicit accuracy assessments at high spatial resolution and across large geographic extents. Based on these computational structures, we efficiently derived continuous, exhaustive surfaces of commonly used agreement metrics and assessed the plausibility (from a geographic perspective) and robustness (to underlying sample size and extreme class imbalance) of these metrics. Moreover, we tested the sensitivity of our results to the choice of assessment support and assessment unit, and applied our framework to multi-temporal built-up surface layers from the GHS-BUILT.

**Implications for analysts conducting accuracy assessments**. We demonstrated that the choice of a suitable agreement measure is critical for conducting meaningful spatially explicit accuracy assessments of binary categorical geospatial data, such as built-up surface products, exemplified herein by the GHSL. The choice of a robust, suitable measure for localized, spatially explicit accuracy assessment is crucial, in particular since class imbalance (and its variability) is a prevalent characteristic of built-up land data. We identified the Intersection-over-Union as the metric yielding most geographically plausible results across the rural-urban continuum, while exhibiting high levels of robustness to underlying assessment support and sample size. Similar results were achieved for the F-measure (Figs. 3 and 4, Table 2). Observed differences between those two measures may be region-specific, and thus, we recommend either the IoU or the less conservative F-measure to be used in localized accuracy assessments of binary spatial data affected by locally varying class imbalance. These findings are robust to the analytical unit, i.e., we can rule out that positional uncertainty in the test or reference data severely biases the thematic accuracy assessment conducted herein, and they are consistent over time (Table 4, Figure A2). Moreover, we identified that some accuracy metrics are heavily sensitive to the underlying sample size (e.g. Kappa, Figure 8c), implying that localized accuracy assessments should be based on spatial support large enough to avoid sample size issues, and small enough to capture desired spatial detail. Accuracy metrics that are sensitive to the sample-size should be avoided if spatial support varies (e.g. if census units are used). Analysts should be aware that the magnitude of most accuracy metrics increases with increasing spatial support (Figure 9).

Furthermore, the work presented herein underlines the well-known drawbacks of reporting overly generalized ("global") accuracy estimates (see e.g., Strahler et al. 2006, Foody 2007, Khatami et al. 2017), and constitutes important methodological knowledge as spatially explicit accuracy assessments are increasingly used in remote-sensing related applications (Morales-Barquero et al. 2019).

**Implications for GHSL practitioners.** The revealed spatial accuracy variations of built-up land in the GHSL will increase awareness of the variability of inherent uncertainty in remote-sensing based settlement data such as the GHSL among data users. We identified high levels of association between accuracy measures and density characteristics of built-up areas, and thus, users can roughly estimate the level of accuracy based on measured built-up density (Table 2, Figure 10). Thus, knowledge, or at least awareness of, fine-grained uncertainty variations in the GHS-BUILT R2018A are essential for an unbiased interpretation of local settlement patterns, and of products derived from GHS-BUILT R2018A such as GHS-POP or GHS-SMOD. Importantly, we provide a refined, spatially explicit view of the increasing GHS-BUILT accuracy trend from rural to urban settings (e.g., Figures 4 and 12, cf. Leyk et al. 2018), resulting in the underestimation of built-up land in rural areas versus overestimation in urban areas. These insights are in line with the findings of related studies using Global Urban Footprint data in Europe (Klotz et al. 2016) and Africa (Mück et al. 2017) and the GHSL in China (Liu et al. 2020). These consistent results suggest that the reported findings are likely to be valid for large parts of North-American settlements, and possibly for comparable landscapes in other regions. Moreover, we shed light on localized accuracy trends over time. We revealed that thematic accuracy has increased considerably, in particular in regions characterized by urban sprawl and densification of built-up areas (Figure 13). However, we also showed that localized densification estimates derived from the GHS-BUILT heavily overestimate the built-up area densification measured by our reference data (Figure 14), calling for GHSL data users to be particularly cautious when using the GHS-BUILT (and its derived products) for local built-up density change assessments.

**Limitations.** The choice of the study area used herein (i.e., the state of Massachusetts) was dictated by data availability and accessibility. Even though this study area is relatively large (>27,000 km²), the observed trends could potentially be biased by relatively homogeneous vegetation and settlement characteristics. As the reported findings are in line with the literature, we are confident that they are valid for large parts of North America. However, some of our results, such as the sensitivity of focal accuracy metrics to the assessment support, as well as the relationships between accuracy and structural characteristics of built-up areas (Figure 6) may be very specific to our study area and could vary considerably if applied to regions of different configurations of built-up surfaces.

It remains to be investigated how these observations differ in regions of different climate and vegetation settings, building materials or settlement configurations. For example, climate-depending frequency of cloud presence or the level of spectral similarity between impervious surfaces and their surrounding natural environments, could affect accuracy trends considerably. In particular, the configuration of rural and peri-urban settlements may vary considerably across geographic regions and could result in different GHS-BUILT accuracy trends across the rural-urban continuum. Moreover, the spatial distribution of training data used for the production of the GHS-BUILT likely affects its accuracy.

While our analysis showed that the effect of positional uncertainty on the observed trends of localized thematic accuracy estimates is largely negligible, it is important to note that part of the disagreement observed in this study is partially due to the different definitions of "settlement" and "built-up area". Settlements encompass buildings, but also impervious surfaces (roads etc.) and small areas of urban greenery (trees, gardens, parks) in direct vicinity of these buildings. This concept of the "generalized" built-up area is implemented in the GHS-BUILT data (Florczyk et al. 2020). Thus, comparing the GHS-BUILT to reference data derived from building footprint data may not be a fair assessment. In particular, the remote-sensing base differentiation between buildings with concrete roofs and paved roads can be difficult due to the similar spectral responses. While the spatial aggregation to 3x3 cell blocks partially mitigates this problem, the incorporation of road network data into the rasterization process to obtain the reference data could further mitigate some of these definitional discrepancies (e.g., Marconcini et al. 2020a). Moreover, temporal inconsistencies between reference construction year and imagery acquisition date of the Landsat data underlying the GHS-BUILT may further affect our accuracy estimates. Such temporal gaps may be caused by heterogeneous levels of currency in the underlying cadastral source data, the vagueness in defining the construction year of a building (i.e., effects of land clearance and construction activities on spectral responses one or two years before a building is finished). While we assume this issue to be of random nature and has only a minor effect on our results, there could be individual clusters of building construction sites, which may affect localized accuracy considerably.

**Future work.** Next steps will apply the proposed framework to larger study areas, and will investigate the potential of using shape and structural properties of built-up areas for predictive uncertainty modeling. Moreover, we will adjust and apply this framework to finer-grained built-up surface layers such as the World Settlement Footprint or novel versions of the GHS-BUILT, as well as to "non-binary" settlement data reporting built-up area fractions or probabilities.

## 3 Appendix

**Table A1. Overview of the accuracy metrics analyzed in this study.**

| Agreement metric | Short name | Alternative name | Accuracy type | Purpose / principle | Criticism |
|---|---|---|---|---|---|
| Precision | - | User's accuracy | Type I error | Measures the commission error | - |
| Recall | - | Producer's accuracy | Type II error | Measures the omission error | - |
| F-measure (F1-score) | - | - | Thematic | Harmonic mean of precision and recall | - |
| Adjusted F-measure | - | - | Thematic | Accounts for class imbalance | - |
| Intersection-over-Union | IoU | Jaccard index, figure of merit | Thematic | Independent from the universe (i.e., from the "true negatives") | - |
| Percentage correctly classified | PCC | Overall accuracy (OA) | Thematic | Takes into account the "true negatives" | Heavily biased in case of dominant negative class |
| Geometric mean | G-mean | - | Thematic | Geometric mean of sensitivity and specificity, accounts for class imbalance | - |
| Cohen's Kappa index | Kappa | - | Thematic | Accounts for chance agreement | Chance agreement is not relevant for classification accuracy assessments, sensitive to marginal probabilities |
| Matthews correlation coefficient | MCC | - | Thematic | Robust to class imbalance | - |
| Normalized mutual information | NMI | - | Thematic | Entropy-based, does not require corresponding class labels | - |
| Absolute error | AE | - | Quantity | Independent from thematic agreement | - |
| Relative error | RE | - | Quantity | Independent from thematic agreement | - |

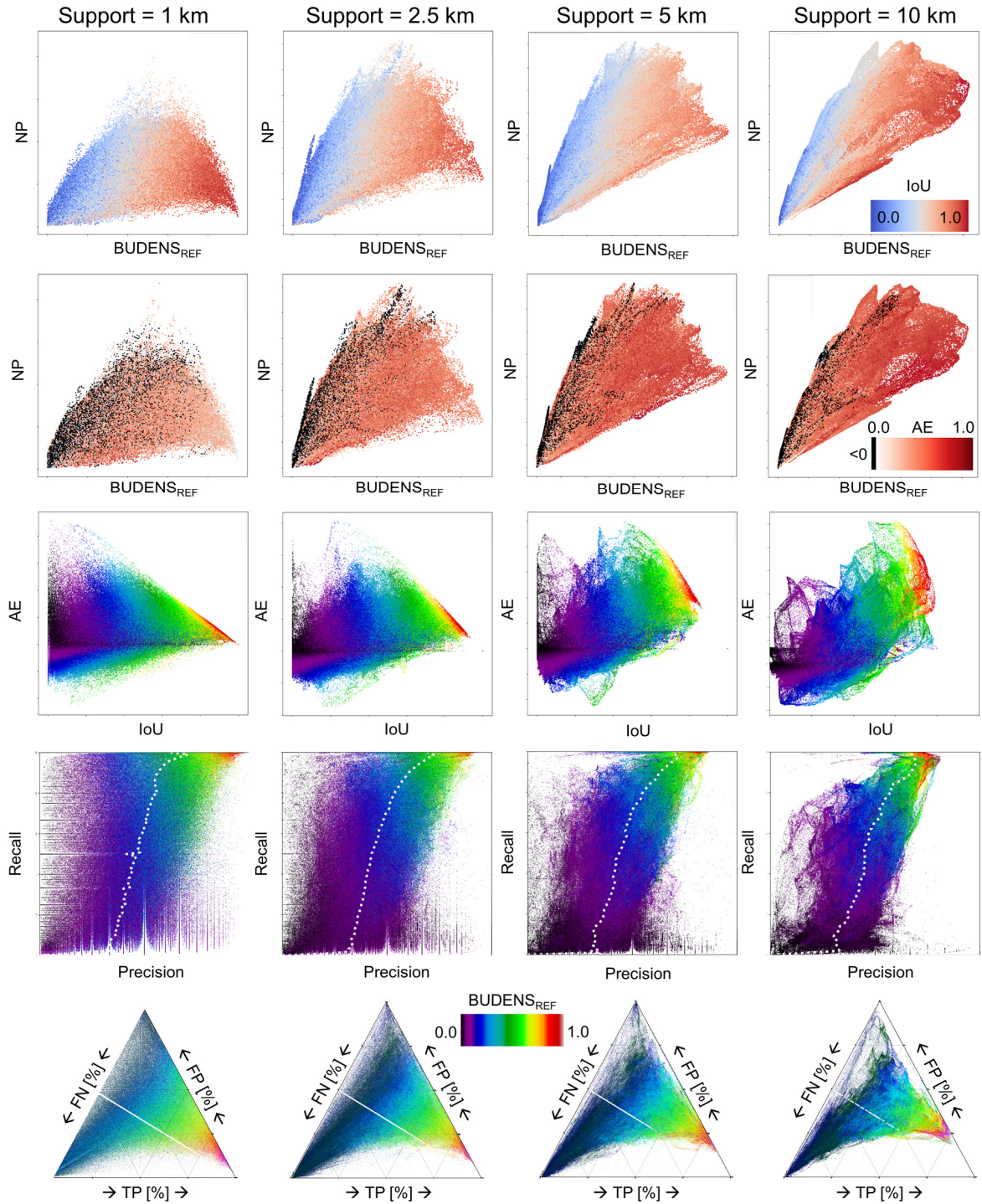

Figure A1. Relationships between accuracy measures, built-up density, number of built-up patches, and agreement categories across levels of assessment support.

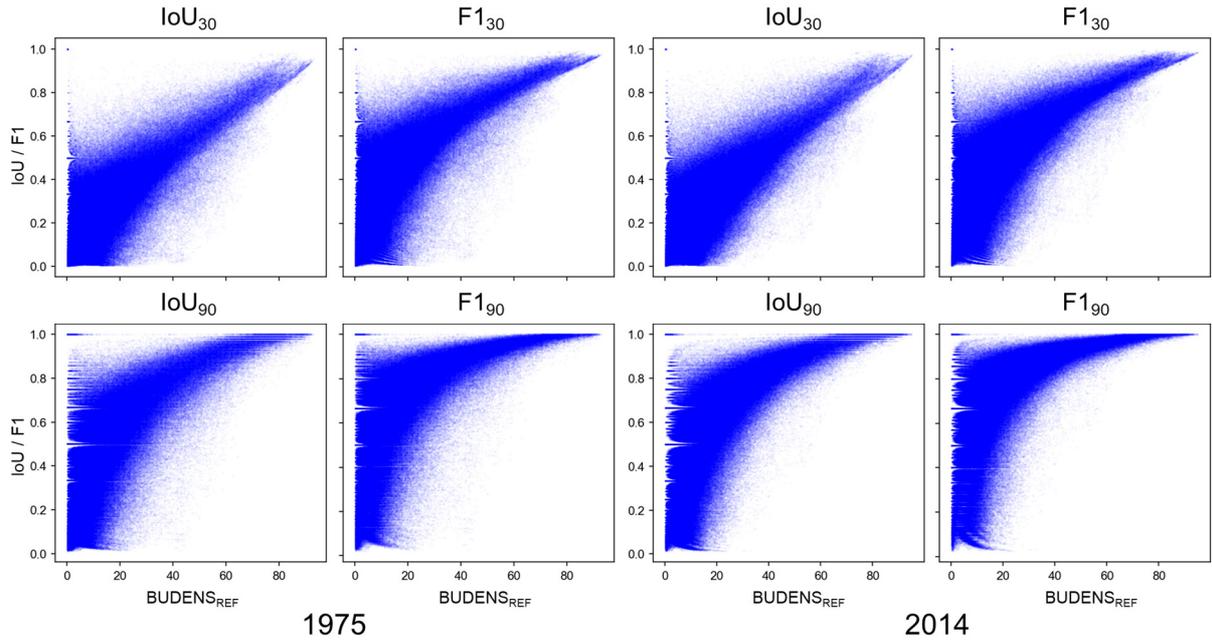

**Figure A2:** Robustness check of trends of IoU and F-measure across the rural-urban continuum over time (i.e., for 1975 and 2014) and for two analytical units (i.e., 30x30m grid cells, and 90x90m blocks).

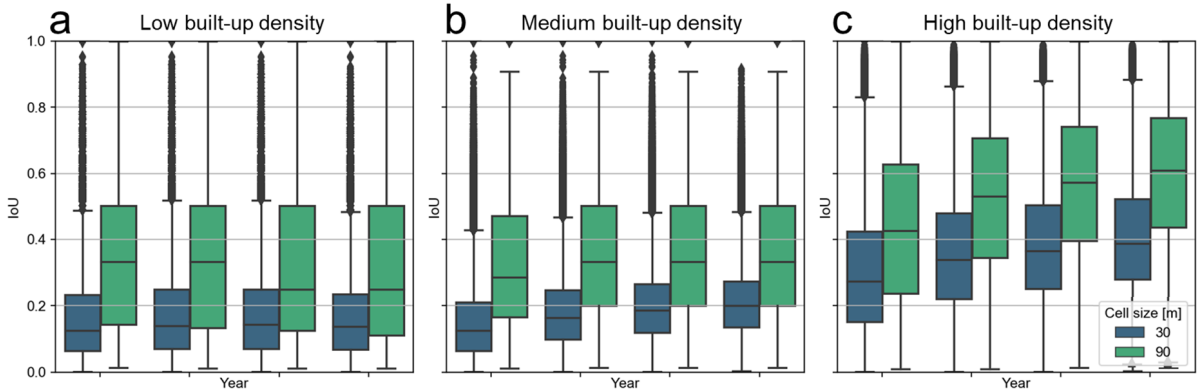

**Figure A3.** Trends of IoU across the four GHSL epochs 1975 – 2014, within strata of reference built-up density, loosely related to (a) rural (0%-2% built-up density), (b) peri-urban (2%-10% built-up density), and (c) urban (>10% built-up density).

**Table A2.** Change statistics of localized IoU from 1975 to 2014, within strata of reference built-up density, for both single cells, and 3x3 cell blocks used as unit for the accuracy assessment. Strata according to Figure A3.

| Stratum | IoU temporal trend | Analytical unit = 30m cells | | | Analytical unit = 3x3 cells | | |
| | | Area proportion [%] | Avg. $\Delta$IoU | Avg. built-up density [%] | Area proportion [%] | Avg. $\Delta$IoU | Avg. built-up density [%] |
|---|---|---|---|---|---|---|---|
| Low-density | Increasing | 47.90 | 0.07 | 1.32 | 45.60 | 0.04 | 1.63 |
| | Decreasing | 52.10 | -0.07 | 1.11 | 54.40 | -0.16 | 1.54 |
| Medium-density | Increasing | 76.53 | 0.09 | 6.36 | 60.18 | 0.13 | 7.04 |
| | Decreasing | 23.47 | -0.05 | 5.57 | 39.82 | -0.13 | 6.54 |
| High-density | Increasing | 92.21 | 0.12 | 28.55 | 87.21 | 0.19 | 48.50 |
| | Decreasing | 7.79 | -0.04 | 20.23 | 12.79 | -0.07 | 34.66 |

# 4    References

Akosa, J. (2017). Predictive accuracy: A misleading performance measure for highly imbalanced data. In Proceedings of the SAS Global Forum (pp. 2-5).

Ariza-López, F. J., Ruiz-Lendínez, J. J., and Ureña-Cámara, M. A. (2018). Influence of Sample Size on Automatic Positional Accuracy Assessment Methods for Urban Areas. ISPRS International Journal of Geo-Information, 7(6), 200.

Belgiu, M., and Drăguţ, L. (2016). Random forest in remote sensing: A review of applications and future directions. ISPRS Journal of Photogrammetry and Remote Sensing, 114, 24-31.

Bujang, M. A., and Baharum, N. (2017). Guidelines of the minimum sample size requirements for Kappa agreement test. Epidemiology, Biostatistics and Public Health, 14(2).

Champagne, C., McNairn, H., Daneshfar, B., and Shang, J. (2014). A bootstrap method for assessing classification accuracy and confidence for agricultural land use mapping in Canada. International Journal of Applied Earth Observation and Geoinformation, 29, 44-52.

Chicco, D., and Jurman, G. (2020). The advantages of the Matthews correlation coefficient (MCC) over F1 score and accuracy in binary classification evaluation. BMC genomics, 21(1), 6.

Cohen, J. (1960). A coefficient of agreement for nominal scales. Educational and Psychological Measurement, 20, 37-46.

Comber, A., Fisher, P., Brunsdon, C., and Khmag, A. (2012). Spatial analysis of remote sensing image classification accuracy. Remote Sensing of Environment, 127, 237-246.

Congalton, R. G. (1988). A comparison of sampling schemes used in generating error matrices for assessing the accuracy of maps generated from remotely sensed data. Photogrammetric engineering and remote sensing (USA).

Congalton, R.G. (1991). A review of assessing the accuracy of classifications of remotely sensed data. Remote sensing of environment 37(1), 35-46.

Congalton, R. G. (2007). Thematic and positional accuracy assessment of digital remotely sensed data. In In: McRoberts, Ronald E.; Reams, Gregory A.; Van Deusen, Paul C.; McWilliams, William H., eds. *Proceedings of the seventh annual forest inventory and analysis symposium*; October 3-6, 2005; Portland, ME. Gen. Tech. Rep. WO-77. Washington, DC: US Department of Agriculture, Forest Service: 149-154. (Vol. 77).

Craig, Belle A, and Jerry L Wahl. 2003. "Cadastral Survey Accuracy Standards." *Surveying and Land Information Science* 63 (2): 87–106.

Corbane, C., Pesaresi, M., Kemper, T., Politis, P., Florczyk, A.J., Syrris, V., Melchiorri, M., Sabo, F. and Soille, P. (2019a). Automated global delineation of human settlements from 40 years of Landsat satellite data archives. *Big Earth Data*, 3(2), pp.140-169.

Corbane, C., Politis, P., Syrris, V., Kempeneers, P., Burger, A., Pesaresi, M., Thomas, K. and Soille, P. (2019b). Automatic image data analytics from a global Sentinel-2 composite for the study of human settlements. In *Proc. Big Data Space* (pp. 89-92).

Corbane, C., Syrris, V., Sabo, F., Politis, P., Melchiorri, M., Pesaresi, M., Soille, P. and Kemper, T. (2021). Convolutional neural networks for global human settlements mapping from Sentinel-2 satellite imagery. *Neural Computing and Applications,* 33(12), pp.6697-6720.

Delgado, R., and Tibau, X. A. (2019). Why Cohen's Kappa should be avoided as performance measure in classification. PloS one, 14(9), e0222916.

Drucker, H. (1997). Improving regressors using boosting techniques. In ICML (Vol. 97, pp. 107-115).

Ehrlich, D., Freire, S., Melchiorri, M., & Kemper, T. (2021). Open and Consistent Geospatial Data on Population Density, Built-Up and Settlements to Analyse Human Presence, Societal Impact and Sustainability: A Review of GHSL Applications. *Sustainability*, 13(14), 7851.

Esch, T., Marconcini, M., Felbier, A., Roth, A., Heldens, W., Huber, M., Schwinger, M., Taubenböck, H., Müller, A. and Dech, S. (2013). Urban footprint processor—Fully automated processing chain generating settlement masks from global data of the TanDEM-X mission. IEEE Geoscience and Remote Sensing Letters 10(6):1617-1621.

ESRI (2020). ArcGIS Python Libraries: ArcPy. Available online: https://www.esri.com/en-us/arcgis/products/arcgis-python-libraries/libraries/arcpy. Last accessed 10 November 2020.

Facebook Connectivity Lab and Center for International Earth Science Information Network - CIESIN - Columbia University (2016). High Resolution Settlement Layer (HRSL). Source imagery for HRSL © 2016 DigitalGlobe. Accessed 23-03-2018.

Fawcett, T. (2006). An introduction to ROC analysis. Pattern recognition letters, 27(8), 861-874.

FGDC (Federal Geographic Data Committee) (1998). Geospatial positioning accuracy standards - Part 3: National standard for spatial data accuracy. Washington, DC: Federal Geographic Data Committee.

Fielding, A.H., and Bell, J.F. (1997). A review of methods for the assessment of prediction errors in conservation presence/absence models. Environmental Conservation, 24(01), 38-49.

Florczyk A.J., Corbane C., Ehrlich D., Freire S., Kemper T., Maffenini L., Melchiorri M., Pesaresi M., Politis P., Schiavina M., Sabo F., Zanchetta L. (2019). GHSL Data Package 2019, EUR 29788 EN, Publications Office of the European Union, Luxembourg, 2019, ISBN 978-92-76-13186-1, doi:10.2760/290498, JRC 117104.

Florczyk, A.J., Melchiorri, M., Zeidler, J., Corbane, C., Schiavina, M., Freire, S., Sabo, F., Politis, P., Esch, T. and Pesaresi, M., 2020. The Generalised Settlement Area: mapping the Earth surface in the vicinity of built-up areas. *International Journal of Digital Earth*, 13(1), pp.45-60.

Foody, G. M. (2002). Status of land cover classification accuracy assessment. Remote sensing of environment, 80(1), 185-201.

Foody, G. M. (2009). Sample size determination for image classification accuracy assessment and comparison. International Journal of Remote Sensing, 30(20), 5273-5291.

Foody, G.M. (2007). Local characterization of thematic classification accuracy through spatially constrained confusion matrices. International Journal of Remote Sensing, 26(6), 1217-1228.

Foody, G. M. (2020). Explaining the unsuitability of the kappa coefficient in the assessment and comparison of the accuracy of thematic maps obtained by image classification. Remote Sensing of Environment, 239, 111630

Forbes, A. D. (1995). Classification-algorithm evaluation: Five performance measures based on confusion matrices. Journal of Clinical Monitoring, 11(3), 189-206.

Freund, Y., and Schapire, R. E. (1997). A decision-theoretic generalization of on-line learning and an application to boosting. Journal of computer and system sciences, 55(1), 119-139.

GDAL/OGR contributors (2020). GDAL/OGR Geospatial Data Abstraction software Library, Open Source Geospatial Foundation, https://gdal.org, Last accessed November 10,2020.

Gong, B., Xu, B., Zhu, Z., Yuan, C., Suen, H.P., Guo, J., Xu, N., Li, W., Zhao, Y. and Yang, J.J.S.B. (2019). Stable classification with limited sample: Transferring a 30-m resolution sample set collected in 2015 to mapping 10-m resolution global land cover in 2017. Sci. Bull, 64, pp.370-373.

Gong, P., Li, X., Wang, J., Bai, Y., Chen, B., Hu, T., Liu, X., Xu, B., Yang, J., Zhang, W. and Zhou, Y. (2020). Annual maps of global artificial impervious area (GAIA) between 1985 and 2018. *Remote Sensing of Environment*, 236, p.111510.

Gu, J., & Congalton, R. G. (2020). Analysis of the impact of positional accuracy when using a single pixel for thematic accuracy assessment. *Remote Sensing*, 12(24), 4093.

Gu, J., & Congalton, R. G. (2021). Analysis of the Impact of Positional Accuracy When Using a Block of Pixels for Thematic Accuracy Assessment. *Geographies*, 1(2), 143-165.

Gwet, K. (2002). Inter-rater reliability: dependency on trait prevalence and marginal homogeneity. Statistical Methods for Inter-Rater Reliability Assessment Series, 2(1), 9.

Hashemian, M. S., Abkar, A. A., and Fatemi, S. B. (2004). Study of sampling methods for accuracy assessment of classified remotely sensed data. In International congress for photogrammetry and remote sensing, 1682-1750.

Herfort, B., Li, H., Fendrich, S., Lautenbach, S., and Zipf, A. (2019). Mapping human settlements with higher accuracy and less volunteer efforts by combining crowdsourcing and deep learning. Remote Sensing, 11(15), 1799.

Jaccard, P. (1902). Gesetze der Pflanzenvertheilung in der alpinen Region. Flora, 90, 349-377.

Khatami, R., Mountrakis, G., and Stehman, S. V. (2017). Mapping per-pixel predicted accuracy of classified remote sensing images. Remote Sensing of Environment, 191, 156-167.

Klotz, M., Kemper, T., Geiß, C., Esch, T., and Taubenböck, H. (2016). How good is the map? A multi-scale cross-comparison framework for global settlement layers: Evidence from Central Europe. Remote Sensing of Environment, 178, 191-212.

Kubat, M. and Matwin, S. (1997). Addressing the curse of imbalanced training sets: one-sided selection. In Proceedings of the 14th International Conference on Machine Learning (ICML), vol. 97, pp. 179-186.

Kyriakidis, P. C., and Dungan, J. L. (2001). A geostatistical approach for mapping thematic classification accuracy and evaluating the impact of inaccurate spatial data on ecological model predictions. Environmental and ecological statistics, 8(4), 311-330.

Leyk, S., and Zimmermann, N.E. (2004). A predictive uncertainty model for field-based survey maps using generalized linear models. International Conference on Geographic Information Science, 191-205.

Leyk, S., and Zimmermann, N. E. (2007). Improving land change detection based on uncertain survey maps using fuzzy sets. Landscape Ecology, 22(2), 257-272.

Leyk, S., and Uhl, J. H. (2018). HISDAC-US, historical settlement data compilation for the conterminous United States over 200 years. Scientific data, 5, 180175.

Leyk, S., Uhl, J. H., Balk, D., and Jones, B. (2018). Assessing the accuracy of multi-temporal built-up land layers across rural-urban trajectories in the United States. Remote sensing of environment, 204, 898-917.

Leyk, S., Gaughan, A.E., Adamo, S.B., de Sherbinin, A., Balk, D., Freire, S., Rose, A., Stevens, F.R., Blankespoor, B., Frye, C. and Comenetz, J. (2019). The spatial allocation of population: A review of large-scale gridded population data products and their fitness for use. Earth System Science Data, 11(3).

Li, M., Ma, L., Blaschke, T., Cheng, L., and Tiede, D. (2016). A systematic comparison of different object-based classification techniques using high spatial resolution imagery in agricultural environments. International Journal of Applied Earth Observation and Geoinformation, 49, 87-98.

Liu, F., Wang, S., Xu, Y., Ying, Q., Yang, F., and Qin, Y. (2020). Accuracy assessment of Global Human Settlement Layer (GHSL) built-up products over China. Plos one, 15(5), e0233164.

Longépé, N., Thibaut, P., Vadaine, R., Poisson, J. C., Guillot, A., Boy, F., ... and Borde, F. (2019). Comparative evaluation of sea ice lead detection based on SAR imagery and altimeter data. IEEE Transactions on Geoscience and Remote Sensing, 57(6), 4050-4061.

Löw, F., Michel, U., Dech, S., and Conrad, C. (2013). Impact of feature selection on the accuracy and spatial uncertainty of per-field crop classification using support vector machines. ISPRS journal of photogrammetry and remote sensing, 85, 102-119.

Luque, A., Carrasco, A., Martín, A., and de las Heras, A. (2019). The impact of class imbalance in classification performance metrics based on the binary confusion matrix. Pattern Recognition, 91, 216-231.

Maratea, A., Petrosino, A., and Manzo, M. (2014). Adjusted F-measure and kernel scaling for imbalanced data learning. Information Sciences, 257, 331-341.

Marc Harper et al. (2015). python-ternary: Ternary Plots in Python. Zenodo. 10.5281/zenodo.594435

Marconcini, M., Metz-Marconcini, A., Üreyen, S., Palacios-Lopez, D., Hanke, W., Bachofer, F., ... and Paganini, M. (2020a). Outlining where humans live, the World Settlement Footprint 2015. Scientific Data, 7(1), 1-14.

Marconcini, M., Gorelick, N., Metz-Marconcini, A., & Esch, T. (2020b). Accurately monitoring urbanization at global scale–the world settlement footprint. In *IOP Conference Series: Earth and Environmental Science* (Vol. 509, No. 1, p. 012036). IOP Publishing.

MassGIS (2016). Office of Geographic Information, Commonwealth of Massachusetts, MassIT, http://www.mass.gov/anf/research-and-tech/it-serv-and-support/application-serv/office-of-geographic-information-massgis/datalayers/. Accessed 18 Aug 2016.

Matthews, B. W. (1975). Comparison of the predicted and observed secondary structure of T4 phage lysozyme. Biochimica et Biophysica Acta (BBA)-Protein Structure, 405(2), 442-451.

McGarigal, K., Cushman, S., Ene, E., (2012). FRAGSTATS v4: spatial pattern analysis program for categorical and continuous maps. Accessible online at: http://www.umass.edu/landeco/research/fragstats/fragstats.html.

McGarigal, K. (2015). FRAGSTATS Help. Accessible online at: https://www.umass.edu/landeco/research/fragstats/documents/fragstats.help.4.2.pdf. Last accessed on 04 October 2020.

Mei, Y., Zhang, J., Zhang, W., and Liu, F. (2019). A Composite Method for Predicting Local Accuracies in Remotely Sensed Land-Cover Change Using Largely Non-Collocated Sample Data. Remote Sensing, 11(23), 2818.

Michie, D., Spiegelhalter D., Taylor, C. (1994). Machine learning, neural and statistical classification, Ellis Horwood.

Mitchell, P. J., Downie, A. L., and Diesing, M. (2018). How good is my map? A tool for semi-automated thematic mapping and spatially explicit confidence assessment. Environmental Modelling & Software, 108, 111-122.

Morales-Barquero, L., Lyons, M. B., Phinn, S. R., and Roelfsema, C. M. (2019). Trends in remote sensing accuracy assessment approaches in the context of natural resources. Remote sensing, 11(19), 2305.

Mück, M., Klotz, M., and Taubenböck, H. (2017, March). Validation of the DLR Global Urban Footprint in rural areas: A case study for Burkina Faso. In 2017 Joint Urban Remote Sensing Event (JURSE) (pp. 1-4). IEEE.

Nelson, J. K., and Brewer, C. A. (2017). Evaluating data stability in aggregation structures across spatial scales: revisiting the modifiable areal unit problem. Cartography and Geographic Information Science, 44(1), 35-50.

Openshaw, S. (1984). The modifiable areal unit problem. Concepts and techniques in modern geography.

Pesaresi, M., Ehrlich, D., Ferri, S., Florczyk, A., Freire, S., Halkia, S., Julea, A., Kemper, T., Soille, P., Syrris, V. (2016). Operating procedure for the production of the Global Human Settlement Layer from Landsat data of the epochs 1975, 1990, 2000, and 2014. JRC Technical Report EUR 27741 EN.

Pesaresi, M., Ehrlich, D., Florczyk, A.J., Freire, S., Julea, A., Kemper, T., Soille, P., Syrris, V.(2015). GHS built-up grid, derived from Landsat, multitemporal (1975, 1990, 2000, 2014). European Commission, Joint Research Centre (JRC) [Dataset] PID: http://data.europa.eu/89h/jrc-ghsl-ghs_built_ldsmt_globe_r2015b

Pesaresi, M., Huadong, G., Blaes, X., Ehrlich, D., Ferri, S., Gueguen, L., ... and Marin-Herrera, M. A. (2013). A global human settlement layer from optical HR/VHR RS data: concept and first results. IEEE Journal of Selected Topics in Applied Earth Observations and Remote Sensing, 6(5), 2102-2131.

Pickard, B., Gray, J., and Meentemeyer, R. (2017). Comparing quantity, allocation and configuration accuracy of multiple land change models. Land, 6(3), 52.

Pontius Jr, R. G., and Millones, M. (2011). Death to Kappa: birth of quantity disagreement and allocation disagreement for accuracy assessment. International Journal of Remote Sensing, 32(15), 4407-4429.

Pontius Jr, R. G., Huffaker, D., and Denman, K. (2004). Useful techniques of validation for spatially explicit land-change models. Ecological Modelling, 179(4), 445-461.

Pontius Jr, R. G., Peethambaram, S., and Castella, J. C. (2011). Comparison of three maps at multiple resolutions: a case study of land change simulation in Cho Don District, Vietnam. Annals of the Association of American Geographers, 101(1), 45-62.

Pontius Jr., R.G. (2002). Statistical methods to partition effects of quantity and location during comparison of categorical maps at multi-ple resolutions. Photogrammetric Engineering and Remote Sensing, 68(10), 1041-1050.

Pontius Jr., R.G. and Cheuk, M.L. (2006). A generalized cross-tabulation matrix to compare soft-classified maps at multiple resolutions. International Journal of Geographical Information Science, 20(1), 1-30.

Pontius Jr., R.G. and Suedmeyer B. (2004). Components of Agreement between categorical maps at multiple resolutions. Remote Sensing and GIS Accuracy Assessment, CRC Press, 233-251.

Pontius, R. G., and Malizia, N. R. (2004). Effect of category aggregation on map comparison. In International Conference on Geographic Information Science (pp. 251-268). Springer, Berlin, Heidelberg.

Pontius, R.G., Boersma, W., Castella, J.C., Clarke, K., de Nijs, T., Dietzel, C., Duan, Z., Fotsing, E., Goldstein, N., Kok, K. and Koomen, E. (2008a). Comparing the input, output, and validation maps for several models of land change. The Annals of Regional Science, 42(1), pp.11-37.

Pontius, R. G., Thontteh, O., and Chen, H. (2008b). Components of information for multiple resolution comparison between maps that share a real variable. Environmental and Ecological Statistics, 15(2), 111-142.

Radoux, J., Waldner, F., and Bogaert, P. (2020). How response designs and class proportions affect the accuracy of validation data. Remote Sensing, 12(2), 257.

Rosenfield, G., and Melley, M. (1980). Applications of statistics to thematic mapping. Photogrammetric Engineering and Remote Sensing, 46, 1287-1294.

Shao, G., Tang, L., and Liao, J. (2019). Overselling overall map accuracy misinforms about research reliability. Landscape Ecology, 34(11), 2487-2492

Sim, J., and Wright, C. C. (2005). The kappa statistic in reliability studies: use, interpretation, and sample size requirements. Physical therapy, 85(3), 257-268.

Smith N. (2000). Scale. In: Johnston RJ, Gregory D, Pratt G, Watts M, editors. The Dictionary of Human Geography. 4th ed. Oxford, UK: Blackwell, pp. 724–27.

Smith, J.H., Stehman, S.V., Wickham, J.D., Yang, L. (2003). Effects of landscape characteristics on land-cover class accuracy. Remote Sensing of Environment, 84(3), 342-349.

Smith, J.H., Wickham, J.D., Stehman, S.V. and Yang, L. (2002). Impacts of patch size and land-cover heterogeneity on thematic image classification accuracy. Photogrammetric Engineering and Remote Sensing 68:65-70

Steele, B.M., Winne, J.C., Redmond, R.L. (1998). Estimation and mapping of misclassification probabilities for thematic land cover maps. Remote Sensing of Environment, 66(2), 192-202.

Stehman, S.V. (2009). Sampling designs for accuracy assessment of land cover. International Journal of Remote Sensing, 30(20), 5243-5272.

Stehman, S.V., and Foody, G. M. (2019). Key issues in rigorous accuracy assessment of land cover products. Remote Sensing of Environment, 231, 111199.

Stehman, S.V., and Wickham, J.D. (2011). Pixels, blocks of pixels, and polygons: Choosing a spatial unit for thematic accuracy assessment. Remote Sensing of Environment, 115(12), 3044-3055.

Stehman, S.V., and Wickham, J. (2020). A guide for evaluating and reporting map data quality: Affirming Shao et al." Overselling overall map accuracy misinforms about research reliability". LANDSCAPE ECOLOGY, 35(6), 1263-1267.

Story, M., and Congalton, R.G. (1986). Accuracy assessment - a users perspective. Photogrammetric Engineering and Remote Sensing, 52(3), 397-399.

Strahler, A.H., Boschetti, L., Foody, G.M., Friedl, M.A., Hansen, M.C., Herold, M., Mayaux, P., Morisette, J.T., Stehman, S.V. and Woodcock, C.E., (2006). Global land cover validation: Recommendations for evaluation and accuracy assessment of global land cover maps. European Communities, Luxembourg, 51(4).

Tsutsumida, N. and Alexis J. Comber (2015). Measures of spatio-temporal accuracy for time series land cover data, International Journal of Applied Earth Observation and Geoinformation, Volume 41, Pages 46-55, https://doi.org/10.1016/j.jag.2015.04.018

U.S. Census Bureau (2017). 2010 Geographic Terms and Concepts. Available online at: https://www.census.gov/geo/reference/terms.html. Last accessed 28 Feb 2017.

Uhl, J.H. and Leyk, S. (2017). Multi-Scale Effects and Sensitivities in Built-up Land Data Accuracy Assessments, Proceedings of International Cartographic Conference 2017, Washington D.C., USA.

Uhl, J. H., and Leyk, S. (2020). Towards a novel backdating strategy for creating built-up land time series data using contemporary spatial constraints. Remote Sensing of Environment, 238, 111197.

Uhl, J. H. & Leyk, S. (2022). MTBF-33: A multi-temporal building footprint dataset for 33 U.S. counties at annual resolution (1900-2015), *Mendeley Data*, V1, doi: 10.17632/w33vbvjtdy

Uhl, J. H., Leyk, S., McShane, C. M., Braswell, A. E., Connor, D. S., & Balk, D. (2021). Fine-grained, spatiotemporal datasets measuring 200 years of land development in the United States. *Earth system science data*, 13(1), 119-153.

Uhl, J. H., Zoraghein, H., Leyk, S., Balk, D., Corbane, C., Syrris, V., and Florczyk, A. J. (2018). Exposing the urban continuum: Implications and cross-comparison from an interdisciplinary perspective. International Journal of Digital Earth.

van Oort, P. A., Bregt, A. K., de Bruin, S., de Wit, A. J., and Stein, A. (2004). Spatial variability in classification accuracy of agricultural crops in the Dutch national land-cover database. International Journal of Geographical Information Science, 18(6), 611-626.

Van Rijsbergen, C. J. (1979). Information Retrieval (2nd ed.). Butterworth-Heinemann.

van Rijsbergen, C.J. (1974). Foundations of evaluation. Journal of Documentation, 30, 365-373.

Vasilakos, C., Kavroudakis, D., and Georganta, A. (2020). Machine learning classification ensemble of multitemporal sentinel-2 images: the case of a mixed mediterranean ecosystem. Remote Sensing, 12(12), 2005.

Waldner, F., Hansen, M. C., Potapov, P. V., Löw, F., Newby, T., Ferreira, S., and Defourny, P. (2017). National-scale cropland mapping based on spectral-temporal features and outdated land cover information. PloS one, 12(8), e0181911.

Waldorf, B., Kim, A. (2018). The Index of Relative Rurality (IRR): US County Data for 2000 and 2010. Purdue University Research Repository. doi:10.4231/R7959FS8

Wardlow, B. D., and Callahan, K. (2014). A multi-scale accuracy assessment of the MODIS irrigated agriculture data-set (MIrAD) for the state of Nebraska, USA. GIScience and remote sensing, 51(5), 575-592.

Webber, J. B. W. (2012). A bi-symmetric log transformation for wide-range data. Measurement Science and Technology, 24(2), 027001.

Wickham, J., Stehman, S. V., and Homer, C. G. (2018). Spatial patterns of the United States National Land Cover Dataset (NLCD) land-cover change thematic accuracy (2001–2011). International journal of remote sensing, 39(6), 1729-1743.

Wickham, J.D., Stehman, S.V., Fry, J.A., Smith, J.H., Homer, C.G. (2010). Thematic accuracy of the NLCD 2001 land cover for the conterminous United States. Remote Sensing of Environment, 114(6), 1286-1296.

Yan, E., Lin, H., Wang, G., and Sun, H. (2014). Multi-scale simulation and accuracy assessment of forest carbon using Landsat and MODIS data. In 2014 Third International Workshop on Earth Observation and Remote Sensing Applications (EORSA) (pp. 195-199). IEEE.

Ye, S., Pontius Jr, R. G., and Rakshit, R. (2018). A review of accuracy assessment for object-based image analysis: From per-pixel to per-polygon approaches. ISPRS Journal of Photogrammetry and Remote Sensing, 141, 137-147.

Zanter, K. 2017. "Landsat Collection 1 Level 1 Product Definition." *United States Geological Survey.*

Zhang, J., and Mei, Y. (2016). Integrating logistic regression and geostatistics for user-oriented and uncertainty-informed accuracy characterization in remotely-sensed land cover change information. ISPRS International Journal of Geo-Information, 5(7), 113.

Zhu, L., Xiao, P., Feng, X., Wang, Z., And Jiang, L. (2013). Multi-scale accuracy assessment of land cover datasets based on histo-variograms. Journal of Remote Sensing, 17(6).